\documentclass{aa} 
\usepackage{txfonts}
\usepackage{longtable}  
\usepackage{rotating}
\usepackage{natbib}
\usepackage{graphicx}
\usepackage{graphics}
\usepackage{psfrag}
\usepackage{amssymb}
\usepackage{multicol}
\bibpunct{(}{)}{;}{a}{}{,}

\def\Teff{\ensuremath{T_{\mathrm{eff}}}}
\def\logg{\ensuremath{\log g}}

\def\vmic{$\upsilon_{\mathrm{mic}}$}

\def\vsini{\ensuremath{{\upsilon}\sin i}}

\def\veq{\ensuremath{{\upsilon}_{\mathrm{eq}}}}
\def\kms{$\mathrm{km\,s}^{-1}$}

\def\ergs{$\mathrm{erg\,s}^{-1}$}

\def\vr{${\upsilon}_{\mathrm{r}}$}
\def\logt{\ensuremath{\log t}}
\def\espa{ESPaDOnS}

\def\nlte{non-LTE}
\def\llm{{\sc LLmodels}}

\def\logl{\ensuremath{\log L/L_{\odot}}}
\def\loglx{\ensuremath{\log L_{\mathrm{X}}}}
\def\loglxlbol{\ensuremath{\log L_{\mathrm{X}}/L_{\mathrm{bol}}}}

\def\M{\ensuremath{M_{\odot}}}
\def\R{\ensuremath{R_{\odot}}}
\def\vald{{\sc VALD}}
\def\synth{{\sc SYNTH3}}
\def\cd{d$^{\rm -1}$}
\def\Msun{$M_{\odot}$}
\def\bonnsai{{\sc BONNSAI}}
\begin{document} 
\title{Two spotted and magnetic early B-type stars in the young open cluster NGC\,2264 discovered by MOST\thanks{Based on data from the MOST satellite, a Canadian Space Agency mission, jointly operated by Microsatellite Systems Canada
Inc. (MSCI), formerly part of Dynacon, Inc., the University of Toronto Institute for Aerospace Studies and the University of British Columbia with the assistance of the University of Vienna.} and \espa\thanks{Based on observations obtained at the Canada-France-Hawaii Telescope (CFHT) which is operated by the National Research Council of Canada, the Institut National des Science de l'Univers of the Centre National de la Recherche Scientifique of France, and the University of Hawaii.}} 
\subtitle{} 
\author{L. Fossati\inst{1}	\and
	K. Zwintz\inst{2}	\and
	N. Castro\inst{1}	\and
	N. Langer\inst{1}	\and
	D. Lorenz\inst{3}	\and
	F.~R.~N. Schneider\inst{1}	\and
	R. Kuschnig\inst{3}	\and
	J.~M. Matthews\inst{4}	\and
	E. Alecian\inst{5,6}	\and
	G.~A. Wade\inst{7}	\and
	T.~G. Barnes\inst{8}	\and
	A.~A. Thoul\inst{9}			   
}
\institute{
	Argelander-Institut f\"ur Astronomie der Universit\"at Bonn, Auf dem 			H\"ugel 71, 53121, Bonn, Germany\\
	\email{lfossati@astro.uni-bonn.de} 
	\and
	Instituut voor Sterrenkunde, K. U. Leuven, Celestijnenlaan 200D, 3001, 			Leuven, Belgium
	\and
	University of Vienna, Department of Astronomy, T\"urkenschanzstrasse 17, 	1180, Vienna, Austria
	\and
	Department of Physics and Astronomy, University of British Columbia, 			6224 Agricultural Road, Vancouver V6T 1Z1, Canada
	\and
	UJF-Grenoble 1 / CNRS-INSU, Institut de Plan\'etologie et 				d'Astrophysique de Grenoble (IPAG) UMR 5274, Grenoble, F-38041, France
	\and
	LESIA-Observatoire de Paris, CNRS, UPMC Univ., Univ. Paris-Diderot, 5 			place Jules Janssen, 92195, Meudon Principal Cedex, France
	\and
	Department of Physics, Royal Military College of Canada, PO Box 17000, 			Stn Forces, Kingston K7K 7B4, Canada
	\and
	The University of Texas at Austin, McDonald Observatory, 82 Mt. Locke 			Rd., McDonald Observatory, Texas, 79734, USA
	\and
	Institut d'Astrophysique et de G\'eophysique, Universit\`e de Li\`ege, 			17 All\'ee du 6 Ao\^ut, 4000, Li\`ege, Belgium
} 
\date{} 
\abstract
{
Star clusters are known as superb tools for understanding stellar evolution. In a quest for understanding the physical origin of magnetism and chemical peculiarity in about 7\% of the massive main-sequence stars, we analysed two of the ten brightest members of the $\sim$10\,Myr old Galactic open cluster NGC\,2264, the early B-dwarfs HD\,47887 and HD\,47777. We find accurate rotation periods of 1.95 and 2.64\,days, respectively, from MOST photometry. We obtained \espa\ spectropolarimetric observations, through which we determined stellar parameters, detailed chemical surface abundances, projected rotational velocities, and the inclination angles of the rotation axis. Because we found only small ($<$5\,\kms) radial velocity variations, most likely caused by spots, we can rule out that HD\,47887 and HD\,47777 are close binaries. Finally, using the least-squares deconvolution technique, we found that both stars possess a large-scale magnetic field with an average longitudinal field strength of about 400\,G. From a simultaneous fit of the stellar parameters we determine the evolutionary masses of HD\,47887 and HD\,47777 to be 9.4$^{+0.6}_{-0.7}$\,\M\ and 7.6$^{+0.5}_{-0.5}$\,\M. Interestingly, HD\,47777 shows a remarkable helium underabundance, typical of helium-weak chemically peculiar stars, while the abundances of HD\,47887 are normal, which might imply that diffusion is operating in the lower mass star but not in the slightly more massive one. Furthermore, we argue that the rather slow rotation, as well as the lack of nitrogen enrichment in both stars, can be consistent with both the fossil and the binary hypothesis for the origin of the magnetic field. However, the presence of two magnetic and apparently single stars near the top of the cluster mass-function may speak in favour of the latter.}
\titlerunning{Two spotted and magnetic early B-type members of NGC\,2264}
\authorrunning{L. Fossati et al.}
\maketitle
\section{Introduction}\label{introduction}
Understanding the evolution of stars is a prerequisite for understanding galaxies and galaxy evolution. In intermediate-mass and massive stars, stellar rotation and binarity \citep{langer2012,maeder2012,sana2012} were identified as playing important roles in stellar evolution, which is currently studied in considerable detail. The role of strong magnetic fields is still poorly understood \citep{langer2012}, even though they have long been known to be present in some intermediate-mass and massive stars.

It has recently been established that the incidence of large-scale magnetic fields in massive and intermediate-mass stars is of the order of 7\% \citep{wade2013}. From the intermediate-mass stars we know that their magnetic fields are either strong ($>$100\,G) or very weak ($<$1\,G), that is, they show a true bi-modality \citep{auriere2007}. While for intermediate-mass stars, magnetic fields give rise to strong chemical peculiarities (Ap, Bp stars), a clear-cut correlation of the presence of a strong magnetic field and abnormal chemical surface abundances in massive stars has not been yet established \citep{martins2012}.

One of the most pressing questions at present is to identify the origin of the strong magnetic fields in intermediate-mass and massive stars. It has been suggested that the magnetic fields might be ``fossil'', that is, inherited from the birth molecular cloud that formed the star, or produced by a pre-main-sequence dynamo \citep[e.g.,][]{moss2001}. Alternatively, the magnetic fields might be produced in the course of a strong binary interaction, for instance, through a stellar binary merger \citep[e.g.,][]{ferrario2009,grunhut2013}. Stars that produce magnetic fields in a pre-main-sequence merger will be difficult to distinguish from those that inherited their field from an earlier stage. However, main-sequence binary interaction will lead to a rejuvenation of the mass gainer (or merger product). Therefore, it may be useful to investigate magnetic stars in star clusters to test these ideas.

\begin{figure}
\includegraphics[width=83mm]{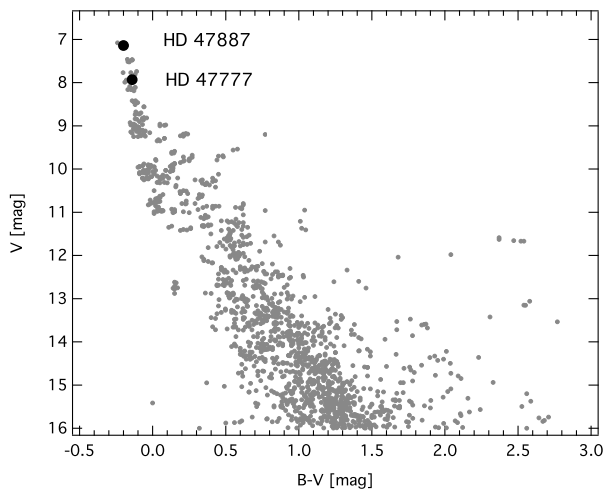}
\caption{Colour-magnitude diagram (CMD) of the open cluster NGC\,2264 in the Johnson $B$ and $V$ filters \citep[grey points, source: WEBDA database of open clusters --][]{webda} showing the positions of HD\,47887 and HD\,47777 \citep{walker1956}.} 
\label{fig:cmd} 
\end{figure}

\begin{figure}
\begin{center}
\includegraphics[width=0.4\textwidth]{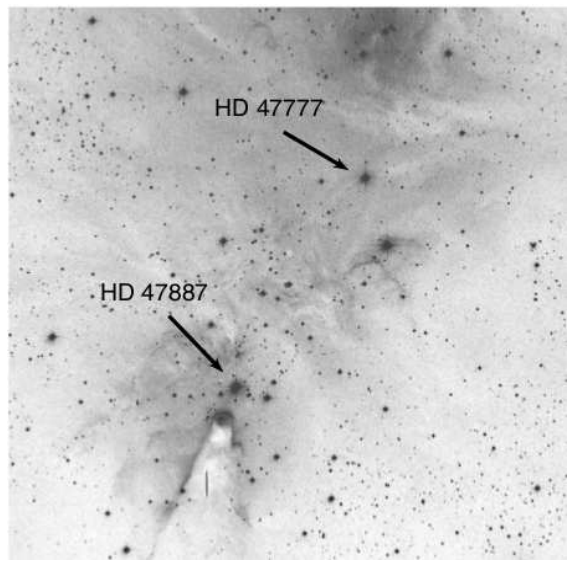}
\caption{DSS image of a 30 x 30 arcminutes$^2$ region showing the positions of HD\,47887 and HD\,47777 in the sky. The ``elephant trunk'' visible at the bottom of the image is probably caused by the intense radiation of the massive cluster-stars falling onto a dense region of the interstellar medium. The geometry of the elephant trunk suggests that HD\,47887 plays an important role in its formation.}
\label{fig:sky} 
\end{center} 
\end{figure}

NGC\,2264 ($\alpha_{2000}$ = 6$^h$\,41$^m$, $\delta_{2000}$  = +9$^\circ$\,53`) was studied frequently in the past using various instruments in different wavelength ranges (from X-ray to radio wavelengths) from space and from the ground \citep[e.g.,][]{sung1997,flaccomio2006,sung2009,zwintz2009}. The cluster is located in the Monoceros OB1 association about 30\,pc above the Galactic plane and has a diameter of $\sim$39\,arcmin. The age of NGC\,2264 can only be determined with a rather large uncertainty as its main sequence consists only of a few massive B-type stars (no main-sequence O-type star is present in the cluster field of view), while stars of later spectral types are still in their pre-main-sequence (PMS) phase \citep[e.g.,][]{zwintz2013a,zwintz2013b}. Different cluster ages have therefore been reported in the literature and range from 3 to 14\,Myr (\logt=6.5--7.15) \citep[e.g.,][]{dambis1999,spassova1985}. Figure~\ref{fig:cmd} shows the colour-magnitude (CMD) diagram of the NGC\,2264 open cluster in Johnson $B$ and $V$ photometry. The stars HD\,47887 (NGC\,2264~178) and HD\,47777 (NGC\,2264~83), two of the brightest, hottest, and most massive stars in the cluster, indicated in Fig.~\ref{fig:cmd} as black circles, are the subject of this work. Their position in the sky is shown in Fig.~\ref{fig:sky}.

With a $V$ magnitude of 7.14 \citep{walker1956}, \object{HD\,47887} is the second-brightest object in the NGC\,2264 open cluster and was classified as B1.5V by \citet{morgan1965}. \citet{flaccomio2000} detected HD\,47887 in the X-rays with the ROSAT HRI instrument, obtaining a maximum X-ray luminosity of \loglx[\ergs]$<$30.26 (\loglxlbol$<-$7.1).

\object{HD\,47777} \citep[$V$ = 7.93 mag][]{walker1956} is a known magnetic early B-type star \citep[B2IV][]{morgan1965}, listed in the \citet{renson2009} catalogue of chemically peculiar stars. For this star, \citet{bychkov2009} reported an average quadratic field modulus of 355\,G on the basis of the analysis of metallic lines, while, making use of the least-squares deconvolution technique \citep[LSD;][]{donati1997}, \citet{alecian2009} presented a firm magnetic field detection. \citet{flaccomio2000} detected HD\,47777 in X-rays with the ROSAT HRI instrument, obtaining an X-ray luminosity of \loglx[\ergs]=30.69  (\loglxlbol$\simeq-$6.3), in agreement with the later measurement of \citet{dahm2007} of \loglx[\ergs]=30.65$\pm$0.04.

\citet{zwintz2009} presented light curves of both HD\,47887 and HD\,47777 obtained with the MOST satellite \citep{walker2003} in 2006, showing a clear rotationally modulated variability. The observed rotational modulation, which is caused by spots, led \citet{zwintz2009} to classify HD\,47887 and HD\,47777 as Bp stars despite of the lack of spectroscopic data.

Following these findings, we obtained additional photometric and spectropolarimetric data to perform a detailed analysis of the two stars. 
\section{Photometric analysis}\label{phot}
\subsection{MOST observations and data reduction}
The Canadian micro-satellite MOST \citep{walker2003} carries a 15 cm Rumak Maksutov telescope that feeds a CCD photometer through a single custom broadband filter in the 3500--7500\,\AA\ wavelength range. The MOST space telescope was launched on June 30, 2003, hence is in its eleventh year of highly successful operation. Three types of photometric data can be supplied simultaneously for multiple targets in a given field: Fabry imaging, direct imaging, and Guide Star Photometry data.

HD\,47887 and HD\,47777 are included in the MOST observing runs on the young open cluster NGC\,2264. Because of the magnitude range (7 $< V <$ 12\,mag) and the large number of targets, the cluster was observed in the open field of the MOST Science CCD in Guide Star Photometry mode. MOST observed the cluster NGC\,2264 twice: the first time in 2006 and the second time in 2011/2012 as part of the CSI2264 (coordinated synoptic investigation of NGC\,2264) project together with the space telescopes CoRoT \citep{baglin2006}, Spitzer \citep{werner2004} and Chandra \citep{weisskopf2002}. During both MOST observing runs two fields of observations were chosen and observed in alternating halves of each 101-minute orbit to increase the number of targets. In 2006 the fields were named field A and field B, in 2011/12 the fields were named field 1 and field 2.
\begin{figure}
\includegraphics[width=83mm]{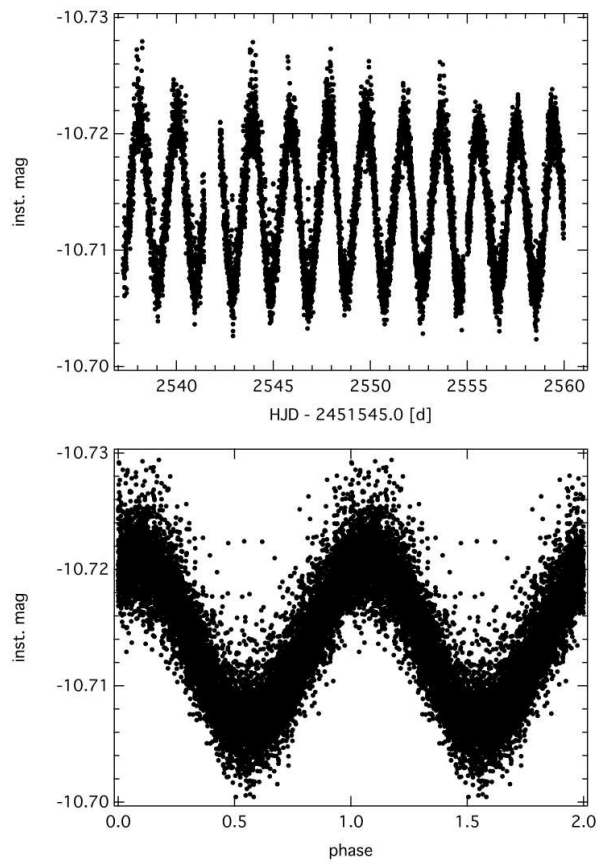}
\caption{Top panel: MOST 2006 light curve of HD\,47887. Bottom panel: light curve phased with a period of 1.949\,days.} 
\label{fig:lc} 
\end{figure}

MOST observed NGC\,2264 from December 7, 2006, to January 3, 2007. On-board exposures were 1.5 seconds long to satisfy the cadence of the Guide Star ACS operations. Sequences of 16 consecutive exposures were stacked on board, sampled about twice per minute. HD\,47887 was included in field A with a total time base of 22.72\,days, while HD\,47777 was observed in field B with a total time base of 22.77\,days. A detailed description of the 2006 MOST data of NGC\,2264 can be found in \citet{zwintz2009}. 
\begin{figure*}
\begin{center}
\includegraphics[width=150mm]{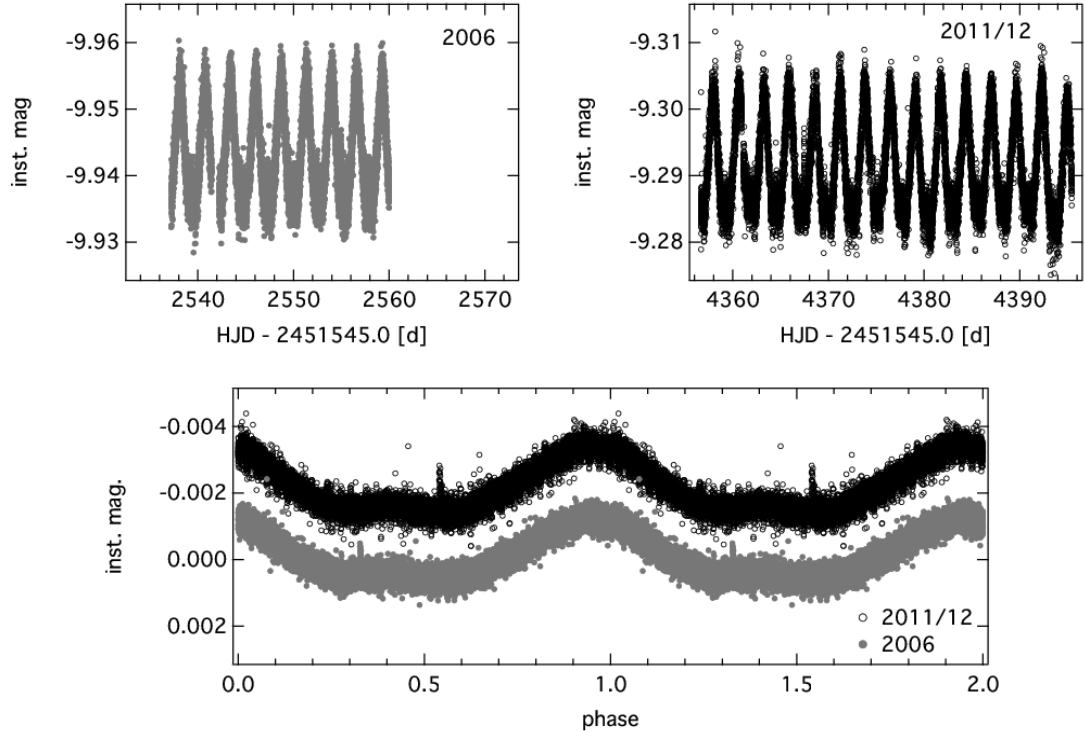}
\caption{MOST light curve and phase plot for HD\,47777: the 2006 data set (grey dots, top-left panel), the 2011/12 data set (black circles, top-right panel), phase plot with a period of 2.641\,days (bottom panel). Note that the 2011/12 data set was shifted for clarity.} 
\label{fig:lc47777} 
\end{center} 
\end{figure*}

During the second MOST run on NGC\,2264 in 2011/2012 the on-board exposures were 3.01\,s long, and the integration time was 51.17\,s as 17 consecutive images were stacked on board. HD\,47777 was observed in field 1 with a time base of 38.77\,days. Because of a slightly different pointing of MOST in 2011/12 compared with 2006, no second light curve of HD\,47887 is available. 

For the reduction of the MOST Guide Star photometry of all stars in the cluster, we used the method developed by \citet{hareter2008} \citep[for more details see][]{zwintz2009}. Consequently, the 2006 light curve of HD\,47887 has 10531 data points (see Fig.~\ref{fig:lc}) with a resulting Nyquist frequency of 1362.23\,\cd; the 2006 data set of HD\,47777 consists of 17698 data points (top-left panel in Fig.~\ref{fig:lc47777}), and the 2011/12 light curve has 16688 data points (top-right panel in Fig.~\ref{fig:lc47777}), corresponding to Nyquist frequencies of 1416.73\,\cd and 721.34\,\cd, respectively. Note that the 2011/12 light curve has fewer data points than the 2006 light curve although the time base is significantly longer. The reason is that in 2011/12 HD\,47777 was observed in the part of the orbit affected by worse stray-light conditions than in 2006, hence more data points had to be discarded in the reduction. 

Table~\ref{mostdata} presents an overview of the properties of the MOST observations. Figures~\ref{fig:lc} and \ref{fig:lc47777} show the MOST light curves of the two stars. 
\begin{table}
\caption{Characteristics of the MOST observations of HD\,47777 and HD\,47887: data set, number of data points used for the analysis, time base, Rayleigh frequency resolution (1/T), Nyquist frequency, and noise level of the residuals (noise$_{\rm res}$). }
\label{mostdata}
\begin{center}
\begin{footnotesize}
\begin{tabular}{llrrcrc}
\hline
\hline
data set & star & points & tbase & 1/T        & f$_{\rm Nyquist}$ & noise$_{\rm res}$ \\
         &      & \#     & [d]   & [d$^{-1}$] & [d$^{-1}$]      & [ppm]               \\
\hline
 2006   & HD\,47777 & 17698 & 22.77 & 0.044 & 1416.73 & 48.5 \\
        & HD\,47887 & 10531 & 22.72 & 0.044 & 1362.23 & 45.9 \\
2011/12 & HD\,47777 & 16688 & 38.77 & 0.026 &  721.34 & 33.9 \\
\hline
\end{tabular}
\end{footnotesize}
\end{center}
\end{table}

%
\subsection{Frequency analysis}
For the frequency analyses, we used the software package Period04 \citep{lenz2005}, which combines Fourier and least-squares algorithms. Frequencies were then prewhitened and considered to be significant if their amplitudes exceeded four times the local noise level in the amplitude spectrum \citep[i.e., 4 S/N;][]{breger1993,kuschnig1997}

We additionally verified the analysis using the SigSpec software \citep{reegen2007}. SigSpec computes significance levels for amplitude spectra of time series with arbitrary time sampling. The probability density function of a given amplitude level is solved analytically and the solution includes dependences on the frequency and phase of the signal.

\begin{figure}
\includegraphics[width=83mm]{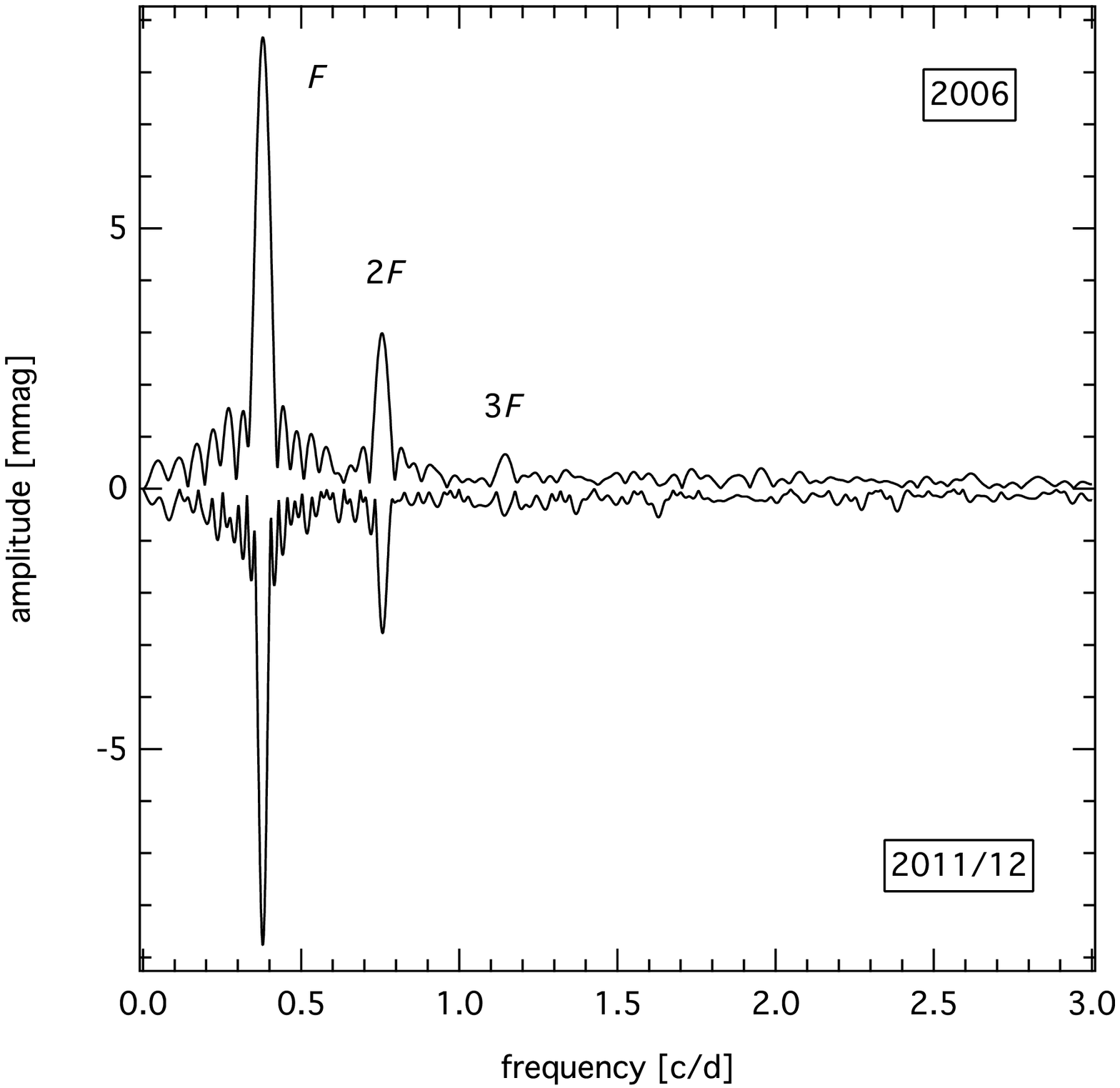}
\caption{Amplitude spectrum of HD\,47777 where $F$ = 0.3787\,\cd\ is marked; twice $F$ and three times $F$ are also marked: 2006 data are pointing upwards, 2011/12 data downwards.} 
\label{fig:amp47777} 
\end{figure}

We found a single intrinsic frequency for the two stars which in both cases was interpreted to be due to rotational modulation of the star. For HD\,47777 we have observations from two epochs that were first analysed separately and the results compared with each other. In both data sets a frequency, $F$, of 0.3787\,\cd (i.e., 4.383\,$\mu$Hz) was identified (see Table~\ref{freqana} and Fig.~\ref{fig:amp47777}), corresponding to a rotation period of 2.641\,days.
Twice $F$ and three times $F$ can also be identified in each of the two data sets separately. The difference in $F$ obtained from the 2006 data and from the 2011/12 data is less than the individual errors on the frequency, hence the frequency does not significantly change within a time span of about five years. The differences in amplitude are also relatively small and amount to 0.111\,mmag. In a second step the two data sets were combined to determine a common solution, obtaining the same frequency $F$ as for the separated analysis. Because frequency and amplitude of the variability do not change over more than five years, the structures causing the rotational modulation of the light curve must be very stable during this period of time. Table~\ref{freqana} lists the frequency and amplitude values for HD\,47777 obtained separately for the two years to illustrate this stability.

\begin{figure}
\includegraphics[width=83mm]{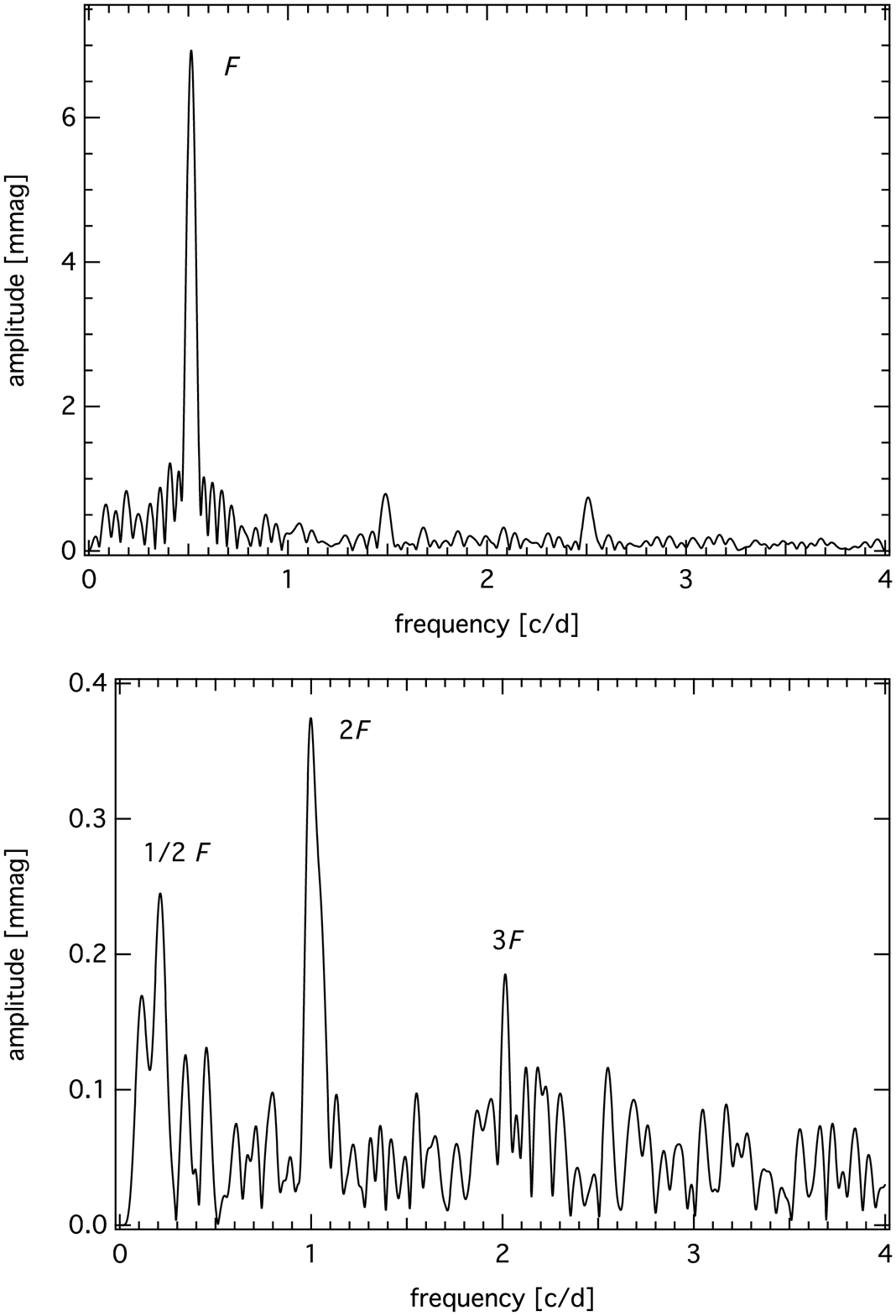}
\caption{Top panel: original amplitude spectrum of HD\,47887 where $F$ = 0.513\,\cd is marked. The two peaks at about 1.5 and 2.5\,\cd\ are one day aliases of $F$. Bottom panel: residual amplitude spectrum where 1/2 $F$, 2 $F$, and 3 $F$ are located at lower amplitudes.} 
\label{fig:amp} 
\end{figure}

For HD\,47887 we found the frequency $F$ at 0.513\,\cd (i.e., 5.94\,$\mu$Hz), as can be seen in the amplitude spectrum shown in the upper panel of Fig.~\ref{fig:amp}. Its multiples, that is, 1/2 $F$, 2 $F$, and 3 $F$, are also present at lower amplitudes (see bottom panel of Fig.~\ref{fig:amp}) and are strong indicators for rotational modulation caused by a non-uniform stellar surface.
\begin{table*}
\caption{Results of the frequency analysis of HD 47777 and HD 47887: frequencies, amplitudes, phases, signal-to-noise values (S/N) and SigSpec significances (sig).}
\label{freqana}
\begin{center}
\begin{tabular}{lllrrcrc}
\hline
\hline
star & data set & \multicolumn{2}{c}{freq} & amp    & period & S/N & sig  \\
     &          & [\cd]     & [$\mu$Hz]    & [mmag] & [days] &     &      \\
\hline
HD\,47777 & 2006    & 0.3786(8) & 4.382(9)  & 8.741   & 2.640  & 8.87   & 3246.02 \\
          & 2011/12 & 0.3787(5) & 4.383(5)  & 8.852   & 2.641  & 12.73  & 3027.30 \\
HD\,47887 & 2006    & 0.513(1)  & 5.94(1)   & 7.060   & 1.949  & 12.19  & 1857.05 \\
\hline
\end{tabular}
\end{center}
\tablefoot{The respective last-digit errors of the frequencies computed according to \citet{kallinger2008} are given in parentheses. The second line for HD\,47777 shows the parameters for $F$ determined from the 2011/12 data set to illustrate its stability.}
\end{table*}

%
\section{Spectroscopic analysis and stellar parameters}\label{spec}
\subsection{Observations and cluster membership}\label{obs}
Spectra of HD\,47887 and of HD\,47777 were obtained on 2012 January 31 (HJD: 2455957.96662) and February 3rd (HJD: 2455960.74013), respectively, with the \espa\ spectropolarimeter of the Canada-France-Hawaii Telescope (CFHT). \espa\ consists of a table-top, cross-dispersed \'echelle spectrograph fed by a double optical fiber directly from a Cassegrain-mounted polarisation analysis module. Both Stokes $I$ and $V$ spectra were obtained throughout the 3700--10400\,\AA\ spectral range at a resolving power of about 65\,000. The spectra were reduced using the Libre-ESpRIT reduction pipeline \citep{donati1997}. The signal-to-noise ratio (S/N) per 1.8\,\kms\ spectral pixel at $\lambda\sim$5000\,\AA\ is 327 and 409, respectively for HD\,47887 and HD\,47777. 

The spectra of the two stars were normalised by fitting a low-order polynomial to carefully selected continuum points. For each star, we determined the radial velocity (\vr) by fitting a Gaussian to the Stokes $I$ LSD profile (see Sect.~\ref{magnetic}). From the \espa\ spectra we obtained \vr=25.2$\pm$0.3\,\kms\ and \vr=26.7$\pm$0.9\,\kms\ for HD\,47887 and HD\,47777, respectively. For HD\,47887, \citet{abt1970} obtained a single radial velocity measurement of 21.6\,\kms. \citet{liu1991} reported a radial velocity of 24.3\,\kms\ and 28.5\,\kms\ for HD\,47887 and HD\,47777, respectively. 

Binary interaction is extremely important in the evolution of massive stars \citep{langer2012,sana2012} and it is therefore important to determine whether HD\,47887 and HD\,47777 are binaries. For HD\,47887 we additionally obtained one high-resolution (R$\sim$60\,000) spectrum in January 2012 with the Robert G. Tull Coud\'e Spectrograph attached to the 2.7 m telescope of Mc\,Donald Observatory and five high-resolution (R$\sim$85\,000) spectra in September 2013 with the HERMES fiber-fed spectrograph \citep{ras11} at the 1.2\,m Mercator telescope on La\,Palma. For HD\,47777 we analysed more high-resolution \espa\ spectra, obtained in December 2007, February 2010, and February 2013 in polarimetric mode, which will be presented in detail by Alecian et al. (in prep.) in the framework of the Magnetism in Massive Stars (MiMeS) project \citep{mimes}. For all spectra we derived \vr\ by fitting a Gaussian to the Stokes $I$ LSD profile (see Sect.~\ref{magnetic}). The average \vr\ value (and its standard deviation) obtained from all analysed spectra is listed in Table~\ref{tab:params_abn}. For both stars we detected weak ($<$5\,\kms) radial velocity variations, typical of spotted stars.

All \vr\ values obtained for the two stars are slightly higher than the average cluster \vr\ of 17.68$\pm$2.26\,\kms, reported by \citet{kharchenko2005} on the basis of 13 stars. In contrast, \citet{liu1991} listed 21 radial velocity measurements for a total of 16 stars in the field of view of NGC\,2264. Except for two stars, which are very likely non-cluster-members or close binaries, the stars observed by \citet{liu1991} indicate a cluster average radial velocity of about 24\,\kms, which agrees well with our measurements.

Despite the possible small difference between the measured radial velocity and the cluster average, the position of the two stars in the sky (see Fig.~\ref{fig:sky}) and the colour-magnitude diagram (compared with that of the other cluster stars), as well as their proper motion \citep{kharchenko2004}, strongly suggest that both stars are members of the NGC\,2264 open cluster.
\subsection{Fundamental parameters and abundance analysis}\label{params}
We determined the projected rotational velocity (\vsini) using the Fourier transform method \citep{simon2007}. The \vsini\ values, listed in Table~\ref{tab:params_abn}, were derived on the basis of the Si\,{\sc iii} line at $\lambda\,\approx$\,4553\,\AA\ and agree well with those reported in the literature by \citet{balona1975} and \citet{vogel1981}. We also searched for macroturbulence broadening, which we modelled as radial-tangential, but found it to be negligible, in agreement with the findings on macroturbulence broadening in massive stars reported by \citet{simon2013}.

Given the higher S/N, we determined the fundamental parameters and chemical abundances using the \espa\ spectra of the two stars obtained in 2012, which were modelled using the {\sc fastwind} \citep[Fast Analysis of STellar atmospheres with WINDs;][]{sr1997,puls2005} stellar atmosphere code. The code enables one to perform \nlte\ calculations and assumes spherical symmetry with an explicit calculation of stellar winds by a $\beta$-like wind velocity law \citep{schaerer1994}. Because the terminal wind velocity cannot be constrained without high-resolution UV spectra \citep{puls2008}, we adopted the values obtained from the empirical calibration given in Eq.~2 of \citet{castro2012} for both stars.

We characterised the spectra on the basis of well-known transitions in the 4000--7000\,\AA\ wavelength range \citep[see][]{crowther2006}. We determined the effective temperature (\Teff) and surface gravity (\logg) by simultaneously fitting the ionisation balance of He (i.e., He\,{\sc i--ii}) and Si (i.e., Si\,{\sc ii--iii--iv}), and the shape of the He\,{\sc i--ii} and H\,{\sc i} Balmer line profiles, searching for the combination of stellar parameters that best reproduced all features taken into account. The parameter determination was based on a large grid of {\sc fastwind} stellar atmosphere models and on a set of routines designed to automatically find the best fit through a $\chi^2$ minimisation algorithm \citep[see][]{lefever2007}. A description of the technique and stellar parameters covered by the adopted grid of stellar atmosphere models is provided by \citet{castro2012}. The determined atmospheric parameters for the two stars are listed in Table~\ref{tab:params_abn}.
\begin{table*}
\caption[ ]{Derived parameters and abundances for HD\,47887 (second column) and HD\,47777 (third column).}
\label{tab:params_abn}
\begin{center}
\begin{tabular}{lcccc}
\hline
\hline
                            &     HD\,47887         &      HD\,47777       & \multicolumn{2}{c}{Sun}     \\
\hline
\Teff\  [K]                 &   24000$\pm$1000      &   22000$\pm$1000     & \multicolumn{2}{c}{5777}    \\
\logg                       &     4.1$\pm$0.1       &     4.2$\pm$0.1      & \multicolumn{2}{c}{4.438}   \\
\vmic\  [\kms]              &       5$\pm$1         &       4$\pm$1        & \multicolumn{2}{c}{0.875}   \\
\vr\  [\kms]                &    22.9$\pm$1.5(7)    &    24.9$\pm$1.3(11)  & \multicolumn{2}{c}{$-$}     \\
\vsini\  [\kms]             &      45               &      60              & \multicolumn{2}{c}{1.2}     \\
Rot. period [d]             &       1.947807        &       2.640426       & \multicolumn{2}{c}{24.47}   \\
\veq\  [\kms]               &     116               &      67              & \multicolumn{2}{c}{2.067}   \\
$i$  [$^\circ$]             &      23               &      64              & \multicolumn{2}{c}{$-$}     \\
\logl\                      &   3.80$\pm$0.15       &   3.42$\pm$0.15      & \multicolumn{2}{c}{0.0}     \\
\smallskip
age$_\mathrm{evol}$ [\logt] & 7.00$^{+0.15}_{-0.18}$&6.92$^{+0.28}_{-0.28}$& \multicolumn{2}{c}{9.66}    \\
\smallskip
M$_\mathrm{evol}$ [\M]      &  9.4$^{+0.6}_{-0.7}$  & 7.6$^{+0.5}_{-0.5}$  & \multicolumn{2}{c}{1.0}     \\
\smallskip
R$_\mathrm{evol}$ [\R]      &  4.4$^{+0.5}_{-0.4}$  & 3.6$^{+0.4}_{-0.3}$  & \multicolumn{2}{c}{1.0}     \\
R$_\mathrm{SED}$ [\R]       &    4.45$\pm$0.50      &    3.50$\pm$0.50     & \multicolumn{2}{c}{1.0}     \\
$<$B$_z>$ [G]               &     373$\pm$48        &     469$\pm$87       & \multicolumn{2}{c}{$-$}     \\
\hline
                &                     &                     &  LTE    & \nlte   \\
He  [dex]       & $-$1.04$\pm$0.10    & $-$1.74$\pm$0.10    & $-$1.04 & $-$1.11 \\
C   [dex]       & $-$4.35$\pm$0.10    & $-$4.15$\pm$0.13    & $-$3.49 & $-$3.61 \\
N   [dex]       & $-$4.36$\pm$0.10    & $-$4.36$\pm$0.13    & $-$4.07 & $-$4.21 \\
O   [dex]       & $-$3.28$\pm$0.10    & $-$3.48$\pm$0.10    & $-$3.17 & $-$3.35 \\
Ne* [dex]       & $-$3.62$\pm$0.08(2) & $-$3.89$\pm$0.05(2) & $-$3.96 & $-$4.11 \\
Mg  [dex]       & $-$5.01$\pm$0.15    & $-$4.81$\pm$0.19    & $-$4.46 & $-$4.44 \\
Al* [dex]       & $-$6.26$\pm$0.08(3) & $-$6.18$\pm$0.10(4) & $-$5.57 & $-$5.59 \\
Si  [dex]       & $-$4.83$\pm$0.10    & $-$4.83$\pm$0.20    & $-$4.49 & $-$4.53 \\
P*  [dex]       &                     & $-$5.93(1)          & $-$6.59 & $-$6.63 \\
S*  [dex]       & $-$5.05(1)          & $-$5.38$\pm$0.11(6) & $-$4.71 & $-$4.92 \\
Ar* [dex]       & $-$5.39(1)          & $-$5.30(1)          & $-$5.52 & $-$5.64 \\
Fe* [dex]       & $-$4.52$\pm$0.32(14)& $-$4.82$\pm$0.11(7) & $-$4.54 & $-$4.54 \\
\hline							
\end{tabular}
\end{center}
\tablefoot{The listed radial velocity \vr\ is the average value obtained from all spectra analysed in this work, and the number in parenthesis indicates the number of spectra for which we determined a \vr\ value. The angle $i$, in degrees, correponds to the inclination angle of the star's rotational axis to the line of sight. R$_\mathrm{evol}$ and R$_\mathrm{SED}$ indicate the stellar radius obtained from the comparison to the evolutionary tracks (see Sect.~\ref{DandC}) and from the fitting of the synthetic fluxes to the observed photometry converted to physical units (see Sect.~\ref{sed}). For comparison, the LTE and \nlte\ solar abundance values obtained by \citet{grevesse1996} and \citet{asplund2009}, respectively, are given in the fourth and fifth columns. The abundances, listed in the second half of the table, are given in $\log (N/N_{\rm tot})$ and the number in parenthesis indicates the number of lines used to determine the element abundance. The abundance of the elements marked with an asterisk was determined assuming LTE on the basis of the final {\sc fastwind} models, as described in Sect.~\ref{params}.}
\end{table*}


Geneva \citep{rufener1966} and Str\"omgren \citep{hauck1997} photometry are available for HD\,47887 and HD\,47777, respectively. Adopting the calibration for the Geneva photometry given by \citet{kunzli1997}, for HD\,47887 we obtained an effective temperature of 25700\,K and a surface gravity of 4.25, which agree, within the error bars, with the values we derived spectroscopically. Table~\ref{tab:photometric_params} lists the fundamental parameters determined from Str\"omgren photometry for HD\,47777, derived by adopting calibrations from \citet{moon1985}, \citet{balona1994}, \citet{napiwotzki1993}, and \citet{castelli1997}. To calibrate the photometry we adopted the TempLogG\,$^{\rm TNG}$ tool described by \citet{kaiser2006}. The effective temperatures derived photometrically and spectroscopically agree well while for \logg\ the photometric values are slightly higher than the spectroscopic ones. Note that \citet{kaiser2006} concluded that \citet{moon1985} provided the most reliable calibration of Str\"omgren photometry for B-type main-sequence stars. In addition, we also note that the \Teff\ value we derived for HD\,47777 agrees well with that reported by \citet{glago1994}.
\begin{table}
\caption[ ]{Effective temperature and surface gravity derived from Str\"omgren photometry \citep{hauck1997} for HD\,47777 and using calibrations from \citet{moon1985}, \citet{balona1994}, \citet{napiwotzki1993}, and \citet{castelli1997}. The last two lines list the \Teff\ and logg\ values derived from the \espa\ spectrum in this work.}
\label{tab:photometric_params}
\begin{center}
\begin{tabular}{lcc}
\hline
\hline
Calibration             & \Teff & \logg \\
                        & [K]   &       \\
\hline
\citet{moon1985}        & 22254 & 4.30 \\
\citet{balona1994}      & 21383 & 4.40 \\
\citet{napiwotzki1993}  & 21847 & 4.55 \\
\citet{castelli1997}    & 20951 & 4.67 \\
\hline
This work               & 22000     & 4.2 \\
                        & $\pm$1000 & $\pm$0.1 \\
\hline							
\end{tabular}
\end{center}
\end{table}


We determined the microturbulence velocity (\vmic) and \nlte\ abundances of C, N, O, Si, and Mg by adopting the set of best-fitting fundamental parameters and using {\sc fastwind} models computed by varying \vmic\ in steps of 1\,\kms\ and the abundance of each considered element in steps of 0.2\,dex. A detailed description of the lines included in the \nlte\ analysis and of the adopted model atoms in the 3900--5000\,\AA\ wavelength range is provided by \citet{castro2012}. Note that the technique has been substantially updated since \citet{castro2012} as follows: the chemical analysis was handled almost in a fully automatic way using a genetic algorithm approach for finding the best composition of models fitting the analysed spectral lines. The \vmic\ and abundance values derived from the \nlte\ analysis are listed in Table~\ref{tab:params_abn}. 

To derive the abundance of the other elements presenting measurable features in the spectra of the two stars, we used the final adopted {\sc fastwind} models to fit synthetic spectra, calculated with \synth\ \citep{synth3}, to the observations. For the spectral synthesis calculation with \synth, we adopted the Vienna Atomic Line Database \citep[\vald;][]{vald1,vald2,vald3} as source for the atomic line parameters. The derived LTE abundances are listed in Table~\ref{tab:params_abn} and are marked with an asterisk (*). Beccause \synth\ assumes LTE, the abundance of the elements obtained in this way has to be taken with caution. The uncertainty on the LTE abundances is the standard deviation from the average abundance and does not account for the uncertainties on the stellar parameters. Following \citet{fossati2009}, the contribution of the uncertainties on \logg\ and \vmic\ on the abundance uncertainties is almost negligible, while, depending on the element, the uncertainty on \Teff\ probably adds another 0.1--0.2\,dex to the abundance uncertainties listed in Table~\ref{tab:params_abn}.

\begin{figure}
\includegraphics[width=83mm]{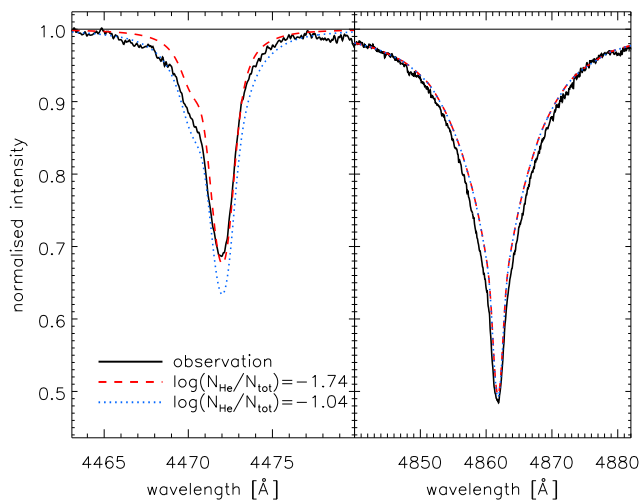}
\caption{Left panel: comparison between the observed He\,{\sc i} line profile at $\sim$4471\,\AA\  (black solid line) of HD\,47777 and \nlte\ synthetic spectra calculated with the solar (blue dotted line) and adopted (red dashed line) He abundance. For both synthetic spectra we adopted the final \Teff\ value of 22000\,K. Right panel: as for the left panel, but for the H$\beta$ line.} 
\label{fig:he} 
\end{figure}

We obtained a solar He abundance for HD\,47887, while HD\,47777 is a He-weak star. The left panel of Fig.~\ref{fig:he} shows the observed HD\,47777 He\,{\sc i} line profile at $\sim$4471\,\AA\ in comparison with \nlte\ synthetic spectra calculated with the solar and adopted He abundance. When fitting the He\,{\sc i} lines with the solar He abundance, we obtained an effective temperature of 16000\,K and a surface gravity \logg\ of 3.4, which disagrees with the photometric parameters and also leads to a considerably poorer fit of the Si\,{\sc ii}/Si\,{\sc iii} ionisation equilibrium, even when the microturbulence velocity is tuned. Because of the He underabundance, we determined the stellar parameters using only the hydrogen Balmer lines and the Si\,{\sc ii}/Si\,{\sc iii} ionisation equilibrium.

For a more consistent comparison, Table~\ref{tab:params_abn} lists the LTE and \nlte\ solar abundances determined by \citet{grevesse1996} and \citet{asplund2009}. For HD\,47887, within the uncertainties most of the abundances derived accounting for \nlte\ are similar to those of the Sun, except for C, Mg, and Si, which are underabundant. For the elements analysed assuming LTE, \nlte\ corrections for early B-type stars are available only for Ne \citep[see e.g.,][]{HH2003,Cunha-Ne,morel08,P08}. \citet{Cunha-Ne} showed that at the temperature of HD\,47887, the Ne \nlte\ correction is about $-$0.5\,dex, bringing our Ne abundance into agreement with the \nlte\ solar abundance value.

For the elements analysed to account for \nlte\ effects, the abundance pattern we obtained for HD\,47777 is remarkably similar (C, Mg, and Si are underabundant, while N and O are solar) to that of HD\,47887, except for a He underabundance of HD\,47777. Assuming the \nlte\ correction given by \citet{Cunha-Ne}, we obtained a Ne underabundance for HD\,47777. For this star, the phosphorus abundance was determined on the basis of just one P\,{\sc i} line for which the oscillator strength was taken from calculations by \citet{H}. Therefore, at least part of the observed overabundance may be caused by uncertainties in the calculated transition probability.
\subsection{High-precision magnetic field detection}\label{magnetic}
\begin{figure*}
\sidecaption
\includegraphics[width=110mm]{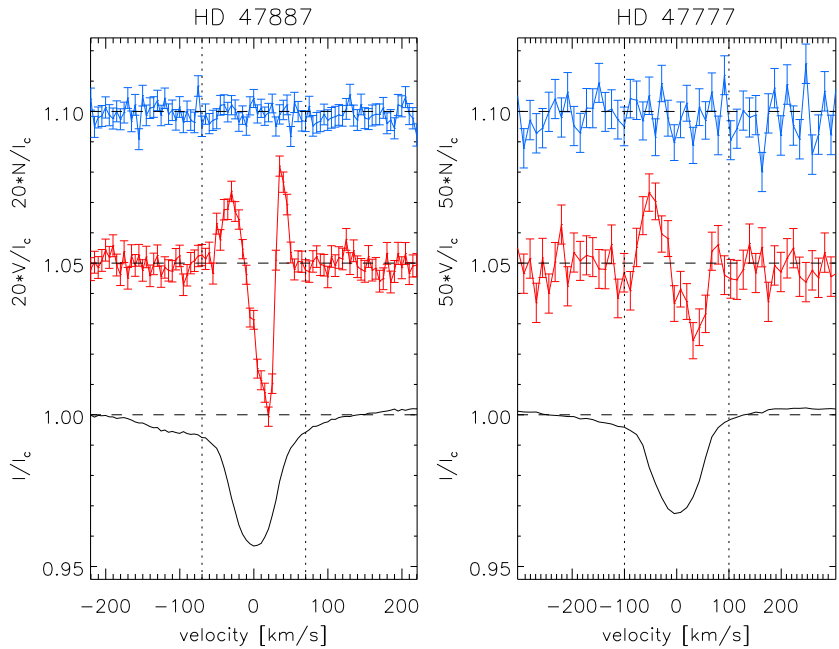}
\caption{$I$, $V$, and $N$ LSD profiles (from bottom to top) obtained for HD\,47887 (left panel) and HD\,47777 (right panel). The $V$ and $N$ profiles were expanded by a factor of 20 for HD\,47887 and 50 for HD\,47777, and were shifted upwards for visualisation purposes. All profiles were shifted to the rest frame. The vertical dotted lines indicate the ranges adopted to calculate the magnetic field.} 
\label{fig:LSDprofiles} 
\end{figure*}

To detect magnetic fields we used LSD \citep{donati1997}, which combines line profiles centred at the position of the individual lines and scaled according to the line strength, wavelength, and sensitivity to a magnetic field. The resulting mean profiles ($I$, $V$, and $N$) were obtained by combining about 200 spectral lines with a strong increase in S/N and therefore sensitivity to polarisation signatures. We computed the LSD profiles of Stokes $I$, $V$, and of the null profile \citep[for a definition of null profile see][]{bagnulo2009} using the methodology and the code described by \citet{kochukhov2010}. We prepared the line mask used by the LSD code separately for the two stars, adopting the stellar parameters and abundances listed in Table~\ref{tab:params_abn}. We extracted the line parameters from \vald. We used all available metallic and He lines stronger than 10\% of the continuum for the two stars, avoiding hydrogen lines and removing the spectral regions affected by telluric lines. The velocity range taken into account to calculate of the average longitudinal magnetic field ($<$B$_z>$) is marked in Fig.~\ref{fig:LSDprofiles}.

Figure~\ref{fig:LSDprofiles} shows the obtained LSD profiles, while Table~\ref{tab:Bfield} gives the results gathered from their analysis. We defined the magnetic field detection making use of the false-alarm probability (FAP): following \citet{donati1992}, we considered a profile with an FAP\,$<\,10^{-5}$ as a definite detection (DD), $10^{-5}\,<$\,FAP\,$\,<10^{-3}$ as a marginal detection (MD), and an FAP\,$>\,10^{-3}$ as a non-detection (ND). To verify that the magnetic field detections are not spurious, we calculated the FAP for the null profile in the same velocity range as used for the magnetic field, obtaining an ND in both cases. We also calculated the FAP for equivalent velocity ranges displaced both redwards and bluewards to sample the continuum in the Stokes $I$ spectrum. The results for both Stokes $V$ and the null profile are given in column 5 of Table~\ref{tab:Bfield}. The FAP obtained in these tests are very close to 1.0, allowing us to conclude that our detections are most likely genuine. In addition, we verified that both Stokes $V$ and the null profile are consistent with the expected noise properties (e.g., Stokes $V$ uncertainties consistent with the standard deviation of the null profile; zero-order moment of Stokes $V$ consistent with zero). Finally, even in rather extreme cases \citep{dela2013}, crosstalk between linear and circular polarisation would not be able to lead to the observed Stokes $V$ signatures, because it would require a strong linear polarisation of the photospheric spectral lines, which is not expected for main-sequence early B-type stars. Moreover, the measured crosstalk of \espa\  (performed periodically since 2010) is negligible\footnote{\tt http://www.cfht.hawaii.edu/Instruments/Spectroscopy\\/Espadons/}.
\begin{table*}
\caption[ ]{Results from the LSD analysis.}
\label{tab:Bfield}
\begin{center}
\begin{tabular}{lccccccc}
\hline
\hline
Star & $<$B$_z>(V)$ & FAP~($V$) & Detection & FAP~($V$) 	   & S/N       & S/N       & \# lines \\
Name & [G]	    & line	& $V$	    & blue cont./red cont. & $I_{LSD}$ & $V_{LSD}$ &         \\
\hline
HD\,47887   &	 373$\pm$48 &         $<$10$^{-15}$& {\bf DD} & 7.8$\times$10$^{-1}$/7.3$\times$10$^{-1}$ & 820 & 5652 & 256 \\
HD\,47777   &	 469$\pm$87 & 2.8$\times$10$^{-12}$& {\bf DD} & 4.7$\times$10$^{-1}$/9.2$\times$10$^{-1}$ &1247 & 7734 & 141 \\
\hline							
Star & $<$B$_z>(N)$ & FAP~($N$) & Detection & FAP~($N$) 	   & S/N     & S/N       & \# lines \\
Name & [G]	    & line	& $N$	    & blue cont./red cont. & $I_{LSD}$ & $V_{LSD}$ &  	\\
\hline
HD\,47887   &	   4$\pm$46 & 7.7$\times$10$^{-1}$ &      ND  & 4.0$\times$10$^{-1}$/9.7$\times$10$^{-1}$ & 820 & 5652 & 256 \\
HD\,47777   &     11$\pm$86 & 2.5$\times$10$^{-1}$ &      ND  & 8.6$\times$10$^{-1}$/2.2$\times$10$^{-2}$ &1247 & 7734 & 141 \\
\hline							
\end{tabular}
\end{center}
\tablefoot{The S/N Stokes $I$ and $V$ is that of the LSD profile. Columns two and three list the average longitudinal magnetic field ($<$B$_z>$) and the FAP for both Stokes $V$ and the $N$ profile. Column four expresses a magnetic field detection on the basis of the FAP, where DD indicates a definite detection (FAP\,$<\,10^{-5}$), MD indicates a marginal detection ($10^{-5}\,<$\,FAP\,$\,<10^{-3}$), and ND incicates a non-detection (FAP\,$>\,10^{-3}$). Column five lists the FAP calculated in the continuum region bluewards and redwards of the spectral line, over a range as broad as that adopted to derive $<$B$_z>$. The last column lists the number of lines used in the line mask.}
\end{table*}

For HD\,47887 the double-reversal shape of the Stokes $V$ LSD profile suggests that the observed stellar surface presents two magnetic poles. On the other hand, for HD\,47777 the shape of the Stokes $V$ LSD profile indicates that the observed stellar surface presents one magnetic pole.
\subsection{Spectral energy distribution}
\label{sed}
%
\begin{figure*}
\sidecaption
\includegraphics[width=120mm]{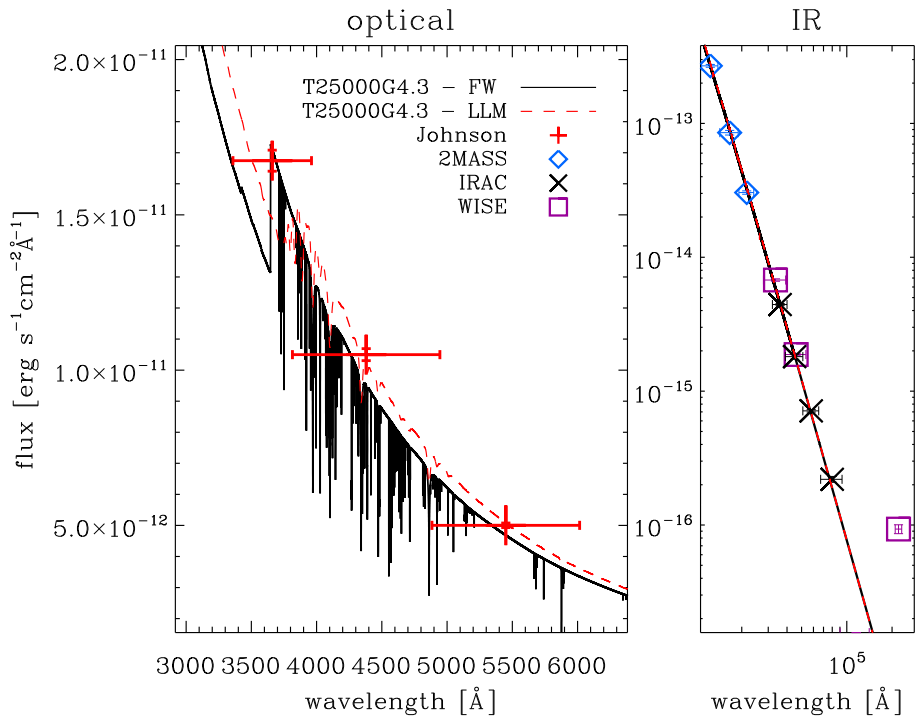}
\caption{Comparison between {\sc fastwind} (full black line) and \llm\ (dashed red line) theoretical fluxes, calculated with the fundamental parameters derived for HD\,47887, with Johnson (red pluses), 2MASS (blue diamonds), Spitzer-IRAC (black crosses), and WISE (purple squares) photometry converted into physical units. The optical and IR spectral regions are shown in the right and left panel, respectively. The synthetic fluxes take into account the cluster distance and reddening given in the text and a stellar radius of 4.45\,\R. The horizontal bars show the full width at half maximum of the photometric filters.} 
\label{fig:sed_hd47887} 
\end{figure*}
\begin{figure*}
\sidecaption
\includegraphics[width=120mm]{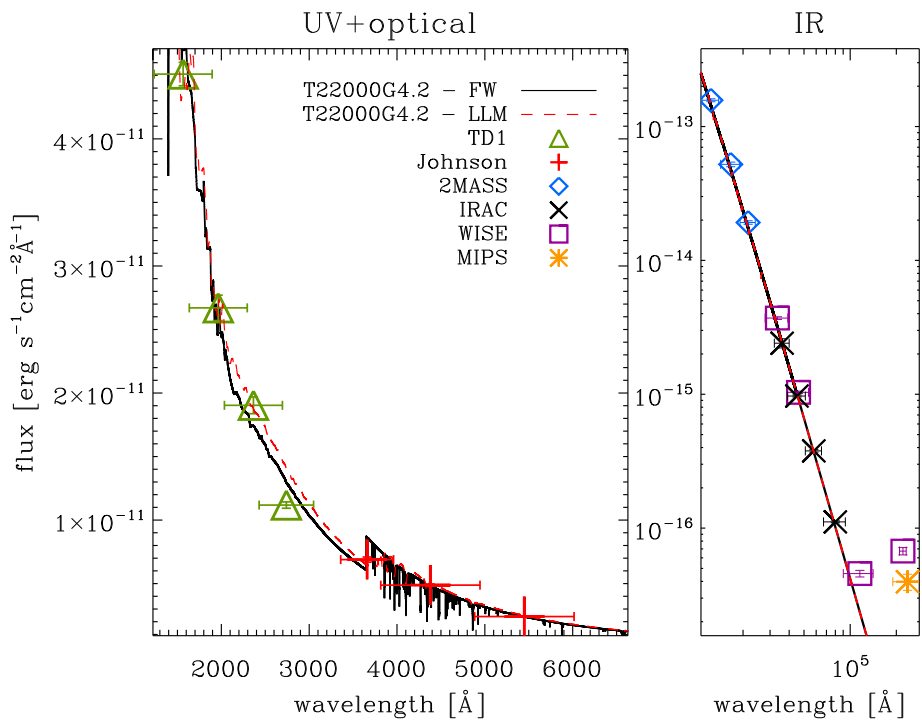}
\caption{Comparison between {\sc fastwind} (full black line) and \llm\ (dashed red line) theoretical fluxes, calculated with the fundamental parameters derived for HD\,47777, with TD1 (green triangles), Johnson (red pluses), 2MASS (blue diamonds), Spitzer-IRAC (black crosses), WISE (purple squares), and Spitzer-MIPS (yellow asterisk) photometry converted into physical units. The UV and optical spectral regions are shown in the left panel, while the right panel shows the IR band. The synthetic fluxes take into account the cluster distance and reddening given in the text and a stellar radius of 3.50\,\R. The horizontal bars show the full width at half maximum of the photometric filters.} 
\label{fig:sed_hd47777} 
\end{figure*}
To examine the quality of the derived atmospheric parameters in more detail and determine the radii of the two stars we compared the synthetic fluxes, calculated with {\sc fastwind}, with flux-calibrated photometry. Figures~\ref{fig:sed_hd47887} and \ref{fig:sed_hd47777} show the comparison of the {\sc fastwind} synthetic fluxes, calculated by adopting the fundamental parameters (and abundances) derived for HD\,47887 and HD\,47777 with the available ultraviolet (UV), optical, and infrared (IR) photometry. We show the Johnson \citep{mermilliod1991}, 2MASS \citep{zacharias2005}, Spitzer \citep[IRAC][]{sung2009}, and WISE \citep{cutri2012} photometry, converted to physical units for the two stars. For the conversion we used the calibrations provided by \citet{bessel1998}, \citet{bliek1996}, \citet{sung2009}, and \citet{wright2010}. For HD\,47777 we also show the available TD1 \citep{thompson1978} photometry and the Spitzer MIPS photometric value at 24\,$\mu$m from \citet{sung2009}, adopting the calibration given by the Spitzer MIPS Instrument handbook\footnote{\tt http://irsa.ipac.caltech.edu/data/SPITZER/docs/mips/
mipsinstrumenthandbook/}. 

Because {\sc fastwind} was not conceived for carrying out analyses in the UV domain and the region across the Balmer jump was calculated without accounting for the high order Balmer lines, we used the \llm\ stellar model atmosphere code \citep{llm} for the two stars to compute LTE synthetic fluxes from the far-UV to the far-infrared wavelength region. \llm\ assumes LTE and plane-parallel geometry and takes into account the individualised abundance pattern for opacity calculations. The \llm\ synthetic fluxes are also shown in Figs.~\ref{fig:sed_hd47887} and \ref{fig:sed_hd47777}.

For the fit performed to determine the stellar radii, we took into account the {\sc fastwind} synthetic fluxes and the TD1 (for HD\,47777), Johnson, and 2MASS photometry \citep[note that we did not consider the TD1 flux value at $\sim$2800\,\AA, which is affected by systematics; see e.g.,][]{morel2013}. We excluded the far-IR photometry from the fit to highlight infrared excess. The adopted cluster distance (743$\pm$80\,pc) and reddening (E(B-V)=0.061$\pm$0.011\,mag) were derived by averaging the results obtained by various photometric studies of this cluster, that is, \citet{gray1965}, \citet{spassova1985}, \citet{lynga1987}, \citet{battinelli1994}, \citet{malysheva1997}, \citet{sung1997}, \citet{dambis1999}, \citet{lotkin2001}, \citet{lata2002}, and \citet{kharchenko2005}. For HD\,47887 we derived a stellar radius of 4.45$\pm$0.50\,\R, while for HD\,47777 we derived a radius of 3.50$\pm$0.50\,\R. Note also the infrared excess shown by the WISE and MIPS photometry, which might be due to a problem in the background subtraction. We corrected the fluxes for the interstellar reddening using the parametrisation by \citet{fitz1999} and assuming a total-to-selective extinction R($V$) of 3.1.

\begin{figure}
\includegraphics[width=83mm]{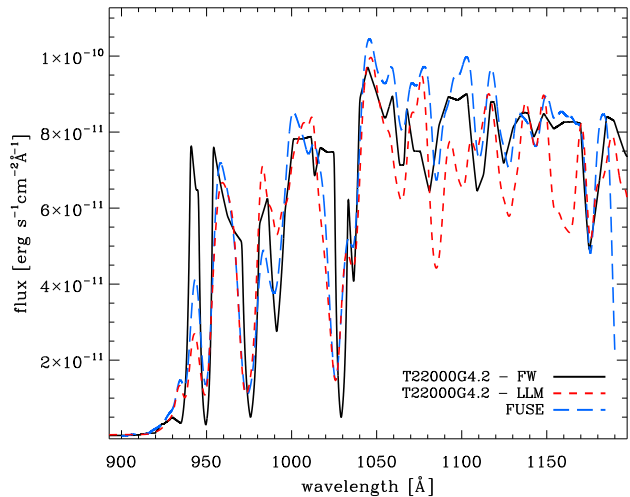}
\caption{Comparison between {\sc fastwind} (full black line) and \llm\ (dashed red line) theoretical fluxes, calculated with the fundamental parameters derived for HD\,47777, with the available flux-calibrated FUSE spectrum (blue long-dashed line). The synthetic fluxes take into account the cluster distance and reddening given in the text and a stellar radius of 3.50\,\R.} 
\label{fig:fuse_hd47777} 
\end{figure}

Figure~\ref{fig:fuse_hd47777} shows a comparison between the {\sc fastwind} and \llm\ synthetic fluxes of HD\,47777 and the available far-UV FUSE spectrum, which we retrieved (reduced and calibrated) from the MAST archive\footnote{\tt http://archive.stsci.edu/}. The synthetic fluxes shown in Fig.~\ref{fig:fuse_hd47777} account for the cluster distance and reddening given above, plus the derived stellar radius of 3.5\,\R. We did not include the FUSE spectrum in the fit of the spectral energy distribution, because it allows one to check the adopted interstellar reddening. Note that in contrast to the other works in which we analysed members of the NGC\,2264 open cluster \citep{zwintz2013a,zwintz2013b}, we decided to adopt a different cluster distance and reddening. Our decision was based on the fact that adopting the reddening of E(B-V)=0.071\,mag, given by \citet{sung1997} and used by \citet{zwintz2013a,zwintz2013b}, we were not able to fit the observed far-UV fluxes, which are instead well reproduced with the lower adopted value of E(B-V)=0.061\,mag. Because the stars belong to a young open cluster, we cannot exclude that the resulting difference in reddening is caused by deviations from the used extinction law, which would be particularly noticeable in the far-UV. 

From the rotational periods derived from the MOST light curves (Sect.~\ref{phot}), the \vsini\ values derived from the high-resolution spectra (Sect.~\ref{params}), and the stellar radii derived from the fit of the spectral energy distribution, we determined the equatorial rotational velocity and the stellar inclination angle. For HD\,47887 we obtained an equatorial rotational velocity (\veq) of 116\,\kms, leading to an inclination angle of 23$^\circ$ (the angle between the star's rotation axis and the line of sight, $i$). For HD\,47777 we derived an equatorial rotational velocity of 67\,\kms\ and an inclination angle of 64$^\circ$. Our results show that HD\,47887 is viewed close to the rotational pole, while HD\,47777 can be considered as viewed almost equator-on. As a consequence, assuming a dipolar magnetic field configuration, the fact that the Stokes $V$ LSD profile of HD\,47887 suggests we observed two poles of the magnetic field implies that the rotation and magnetic axis have different inclination angles, as found for most magnetic stars. Nevertheless, it is possible that HD\,47887 has a complex magnetic field geometry, as shown by other magnetic early B-type stars, such as $\tau$\,Sco \citep{donati2006}, HD\,37776 \citep{kochukhov2011}.
\subsection{Hertzsprung--Russell diagram}\label{hrdiagram}
\begin{figure}
\includegraphics[width=83mm]{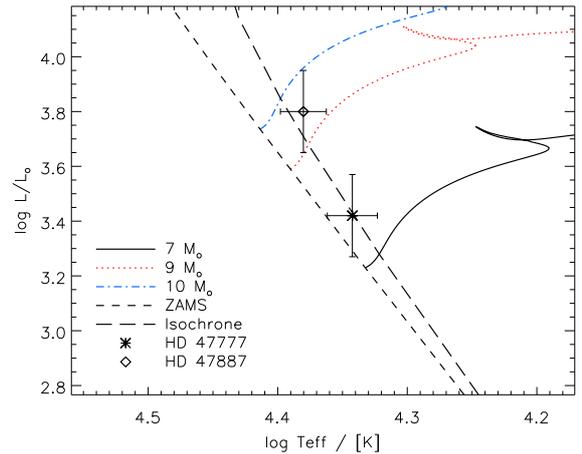}
\caption{Position of HD\,47887 (diamond) and HD\,47777 (asterisk) in the Hertzsprung--Russell (HR) diagram in comparison with Milky Way composition evolutionary tracks for 7\,\M\ (black solid line), 9\,\M\ (red dotted line), and 10\,\M\ (blue dash-dotted line) stars computed by \citet{brott}. The black short-dashed line shows the zero-age main sequence (ZAMS), while the black long-dashed line shows the isochrone calculated with the adopted cluster age of \logt=6.89.} 
\label{fig:hr} 
\end{figure}

Figure~\ref{fig:hr} shows the position of HD\,47887 and HD\,47777 in the Hertzsprung--Russell (HR) diagram in comparison with Milky Way composition evolutionary tracks for 7\,\Msun, 9\,\Msun, and 10\,\Msun\ stars \citep{brott}. We calculated the luminosities of the two stars on the basis of the Johnson $V$-band magnitude \citep{walker1956} and of the distance and reddening adopted in Sect.~\ref{sed}. We used the bolometric correction provided by the {\sc fastwind} model, which agrees well with that obtained with the calibration given by \citet{balona1994}. Luminosities are listed in Table~\ref{tab:params_abn} and their uncertainties take into account the uncertainty in the distance and the bolometric correction, the latter assumed to be 0.1\,mag.

We determined the masses, radii, and ages of the two stars using the \bonnsai\ code \citep[][Schneider et al., in prep.]{schneider2013a} and the Milky Way stellar evolution models of \citet{brott}. \bonnsai\ computes the posterior probability distribution, $p(\vec{m}|\vec{d})$, of stellar model parameters, $\vec{m}$, given observational data, $\vec{d}$, using Bayes' theorem:
\begin{equation}
p(\vec{m}|\vec{d}) \propto p(\vec{d}|\vec{m}) p(\vec{m})\,, \label{eq:bayes-theorem}
\end{equation}
where $p(\vec{d}|\vec{m})$ is the likelihood and $p(\vec{m})$ the prior. We adopted a Gaussian likelihood, used a standard Salpeter initial mass function \citep{salpeter1955} as prior of the initial mass as well as flat priors for the two remaining stellar model parameters, stellar age, and initial equatorial rotational velocity (the initial chemical composition of stars is not a free parameter because we used stellar models with a fixed chemical composition). Adopting flat priors means that all stellar ages and initial equatorial rotational velocities are a priori equally probable. The advantages of this method for our application are that (i) all available observables are simultaneously matched to stellar models, taking the observed uncertainties and a priori knowledge into account, and (ii) robust uncertainties for the stellar parameters are derived from the posterior probability distribution.

We simultaneously match the observational data $\vec{d}=(\logl,\,\Teff,\,\logg,\,\veq)$ (see Table~\ref{tab:params_abn}) to stellar models to compute the posterior probability distribution of the stellar parameters. We adopted $1\sigma$ uncertainties of $0.15\,\mathrm{dex}$ on \logl, 1000\,K on \Teff, $0.1\,\mathrm{dex}$ on \logg\ and 20\,\kms\ on \veq. The stellar models reproduce the set of observables well, that is, each observable is not farther away from the best-matching stellar model than $\sim0.2\sigma$.

For HD\,47887 and HD\,47777 we obtained initial stellar masses of 9.4$^{+0.6}_{-0.7}$\,\M\ and 7.6$^{+0.5}_{-0.5}$\,\M, and stellar radii of 4.4$^{+0.5}_{-0.4}$\,\R\ and 3.6$^{+0.4}_{-0.3}$\,\R, respectively. The most likely initial equatorial rotational velocities are 120$^{+19}_{-23}$\,\kms\ and 70$^{+19}_{-24}$\,\kms, respectively. The given uncertainties are $1\sigma$ confidence intervals. Mass loss can be considered to be negligible because we obtained a mass loss of about 3$\times$10$^{-10}$\,\M/yr for both stars.

The stellar radii obtained using the evolutionary tracks agree well with those we derived from the SED analysis (see Sect.~\ref{sed}). Note that the use of the stellar radii obtained from the evolutionary tracks changes the equatorial rotational velocity by only about 2\,\kms\ for both stars. \citet{hohle2010} used evolutionary tracks from three different authors to determine the mass of HD\,47777. They obtained an average value of 6.31$\pm$3.54\,\M, which, given the very large uncertainty, agrees with our estimated mass within one sigma. Nevertheless, the slightly lower value is most likely due to the lower effective temperature of 18700\,K adopted by \citet{hohle2010}, which was based upon the B3V spectral classification.

We also determined the age for the two stars, obtaining \logt=7.00$^{+0.15}_{-0.18}$ and \logt=6.92$^{+0.28}_{-0.28}$ for HD\,47887 and HD\,47777, respectively. These values agree well with the cluster age of \logt=6.89$\pm$0.26, which we derived in the same way (i.e., averaging) as for the cluster distance and reddening (see Sect.~\ref{sed}). This agreement is also evident in the positions of the two stars on the HR diagram with an isochrone calculated for the adopted cluster age (see Fig.~\ref{fig:hr}). 
\section{Discussion and conclusions}\label{DandC}
We have derived the physical properties of HD\,47887 and HD\,47777. The two stars are found to be core-hydrogen burning stars with evolutionary masses of 9.4\,\M\ and 7.6\,\M\ and possess strong magnetic fields of several hundred Gauss. Our photometric results imply that both stars show brightness inhomogeneities at their surfaces, which allowed us to derive their rotation periods of 1.9\,d and 2.6\,d, with high accuracy. We conclude that both stars are rather slow rotators, with equatorial rotational velocities of 116\,\kms\ and 67\,\kms. 

Despite their similarity, there are distinct differences between the two stars. Our results are consistent with a simple, dipolar magnetic field structure for HD\,47777, while HD\,47887 appears to show two magnetic poles at its visible surface, possibly suggesting a more complex magnetic field geometry, perhaps similar to $\tau$\,Sco \citep{donati2006}. Furthermore, our atmosphere analysis indicates that the surface abundances of HD\,47887 and HD\,47777 are rather similar, but HD\,47777 shows a strong He underabundance. Because the two stars are members of the same cluster, such a difference is probably not caused by different initial conditions, but by evolutionary processes. Because the low helium abundance of He-weak chemically peculiar stars is understood to be an effect of diffusion in the stellar envelope \citep[see e.g.,][]{michaud1970,michaud1979}, our results imply that this effect can occur in stars of $\sim$7.6\,\M, but not at the slightly higher mass of 9.4\,\M. This could either be due to stronger subsurface convection at higher mass \citep{cantiello2009}, or a strengthening of stellar-wind-induced mass loss \citep[cf.,][]{vick2010}. (Note that the slightly higher equatorial rotational velocity and weaker magnetic field of the more massive star might also play a role.) In either case, the two stars analysed here are ideal candidates to test these conjectures.

The two stars we analysed may also shed light on the question of the origin of magnetic fields. One variant of the fossil-field hypothesis proposes that the magnetic fields in intermediate-mass and massive main-sequence stars are surviving stable remnants of the interstellar magnetic fields that thread the molecular clouds from which the stars formed \citep{donati2009}. Since then the field has no other consequences than lowering the angular momentum of the forming stars, none of the derived properties of the two stars contradicts a fossil origin of the magnetic field.

Alternatively, the magnetic fields of the two stars could have been produced during a strong binary interaction, for instance, by a stellar merger \citep{langer2013}. While a merger product is expected to rotate rapidly at first \citep{demink2013}, a strong magnetic field generated during this process may induce substantial angular momentum loss and consequently slow rotation
\citep{langer2012}. Furthermore, \citet{gleb2013} predicted that while massive
merger stars in general often show a surface nitrogen enrichment, their results
also imply that in the mass range considered here merger stars may retain
their abundances unaltered. Consequently, while the observed spin and surface
abundances of HD\,47887 and HD\,47777 do not clearly support the binary
hypothesis, neither do they dispute it.

A merging during main-sequence evolution leads to a rejuvenation of the merger product and transforms it into a blue straggler, whereas a pre-main-sequence merger caused by tidal effects from circumbinary material \citep{korn2012} would lead to a main-sequence star of normal age. Blue stragglers would naturally populate the high-mass end of the mass distribution of cluster stars \citep{schneider2013b}. While Figs~\ref{fig:cmd} and \ref{fig:hr} show no clear evidence for HD\,47887 and HD\,47777 to be blue straggler stars, the upper main-sequence of NGC\,2264 is populated only very sparsely. The fact that at least two of the ten brightest stars of NGC\,2264 are magnetic, whereas the incidence of magnetic fields in early B-type stars in general is about 7\%, might be interpreted as support of the binary hypothesis.

We conclude that it is highly worthwhile to investigate the remaining massive stars in the NGC\,2264 open cluster in some detail. Their stellar parameters may allow us to determine the age of this cluster more accurately, and to identify the blue straggler nature of any of the cluster stars. On the other hand, any detection of a magnetic field in the remaining about ten most massive cluster members would be spectacular, because it would render NGC\,2264 the cluster with the highest concentration of magnetic main-sequence massive stars, together with the Orion Nebula Cluster, where three out of nine main-sequence massive stars are magnetic.
\section*{Acknowledgments}
The research leading to these results has received funding from the European Research Council under the European Community's Seventh Framework Programme (FP7/2007 -- 2013)/ERC grant agreement No 227224 (PROSPERITY). This research is (partially) funded by the Research Council of the K.U. Leuven under grant agreement GOA/2013/012. KZ received a Pegasus Marie Curie Fellowship of the Research Foundation Flanders (FWO) during part of this work. GAW acknowledges support from the Natural Sciences and Engineering Research Council of Canada (NSERC). AAT is a Chercheur Qualifi\'e FNRS. Spectroscopic data were obtained with the 2.7-m telescope at Mc Donald Observatory, Texas, US. Based on observations made with the Mercator Telescope, operated on the island of La Palma by the Flemish Community, at the Spanish Observatorio del Roque de los Muchachos of the Instituto de Astrof{\'i}sica de Canarias. Based on observations obtained with the HERMES spectrograph, which is supported by the Funds for Scientific Research of Flanders (FWO), Belgium, the Research Council of K.U. Leuven, Belgium, the Fonds National Recherches Scientific (FNRS), Belgium, the Royal Observatory of Belgium, the Observatoire de Gen\`eve, Switzerland and the Th{\"u}ringer Landessternwarte Tautenburg, Germany. Use was made of the WEBDA database of open clusters developed and maintained by E. Paunzen, C. St\"utz and J. Janik at the Department of Theoretical Physics and Astrophysics of the Masaryk University, Brno, Czech Republic. We would also like to thank P. Beck and P. Degroote for fruitful discussions, A. Moffat for useful comments on the manuscript, and E. Paunzen and I. M\"uller for their support in the preparation of this work.


\begin{thebibliography}{}
\bibitem[Abt(1970)]{abt1970}
	Abt, H.~A. 1970, ApJS, 19, 387
\bibitem[Ahumada \& Lapasset(1995)]{ahumada1995}
	Ahumada, J. \& Lapasset, E. 1995, A\&AS, 109, 375
\bibitem[Alecian et al.(2009)]{alecian2009}
	Alecian, E., Catala, C., Wade, G.~A., et al. 2009, Magnetism in Herbig 			Ae/Be stars and the link to the Ap/Bp stars, EAS Publications Series, 			eds C. Neiner \& J.-P. Zahn, 39, 121
\bibitem[Asplund et al.(2009)]{asplund2009}
	Asplund, M., Grevesse, N., Sauval, A.~J. \& Scott, P. 2009, ARA\&A, 47, 		481
\bibitem[Auriere et al.(2007)]{auriere2007} 
	Auri{\`e}re, M., Wade, G.~A., Silvester, J., et al. 2007, A\&A, 475, 			1053
\bibitem[Baglin(2006)]{baglin2006} 
	Baglin, A. 2006, The CoRoT mission, pre-launch status, stellar 				seismology and planet finding (M. Fridlund, A. Baglin, J. Lochard and L. 	Conroy eds, ESA SP-1306, ESA Publication Division, Noordwijk, The 			Netherlands)
\bibitem[Bagnulo et al.(2009)]{bagnulo2009}
	Bagnulo, S., Landolfi, M., Landstreet, J.~D., Landi Degl'Innocenti, E., 		Fossati, L. \& Sterzik, M. 2006, PASP, 121, 993
\bibitem[Balona(1975)]{balona1975}
	Balona, L.~A. 1975, MNRAS, 78, 51
\bibitem[Balona(1994)]{balona1994}
	Balona, L.~A. 1994, MNRAS, 268, 119
\bibitem[Battinelli et al.(1994)]{battinelli1994}
	Battinelli, P., Brandimarti, A. \& Capuzzo-Dolcetta, R. 1994, A\&AS, 			104, 379
\bibitem[Bessel et al.(1998)]{bessel1998}
	Bessel, M. S., Castelli, F. \& Plez, B. 1998, A\&A, 333, 231
\bibitem[Brott et al.(2011)]{brott}
	Brott, I., de Mink, S.~E., Cantiello, M., et al. 2011, A\&A, 530, A115
\bibitem[Breger(1993)]{breger1993} 
	Breger, M. 1993, A\&A, 271, 482
\bibitem[Bychkov et al.(2009)]{bychkov2009}
	Bychkov, V.~D., Bychkova, L.~V. \& Madej, J. 2009, MNRAS, 394, 1338
\bibitem[Cantiello et al.(2009)]{cantiello2009}
	Cantiello, M., Langer, N., Brott, I., et al. 2009, A\&A, 499, 279
\bibitem[Castelli et al.(1997)]{castelli1997}
	Castelli, F., Gratton, R.~G. \& Kurucz, R.~L. 1997, A\&A, 318, 841
\bibitem[Castro et al.(2012)]{castro2012}
	Castro, N., Urbaneja, M.~A., Herrero, A., et al. 2012, A\&A, 542, A79
\bibitem[Crowther et al.(2006)]{crowther2006}
	Crowther, P.~A., Lennon, D.~J. \& Walborn, N.~R. 2006, A\&A, 446, 279
\bibitem[Cunha et al.(2006)]{Cunha-Ne}
 	Cunha, K., Hubeny, I., \& Lanz, T. 2006, ApJ, 647, L143
\bibitem[Cutri et al.(2012)]{cutri2012}
	Cutri, R. M., et al. 2012, WISE All-Sky Data Release, VizieR On-line 			Data Catalog: II/311
\bibitem[Dahm et al.(2007)]{dahm2007}
	Dahm, S.~E., Simon, T., Proszkow, E.~M. \& Patten, B.~M. 2007, AJ, 134, 		999
\bibitem[Dambis(1999)]{dambis1999}
	Dambis A. K. 1999, Astronomy Letters, 25, 7
\bibitem[de la Chevroti{\`e}re et al.(2013)]{dela2013}
	de la Chevroti{\`e}re, A., St-Louis, N., Moffat, A.~F.~J. \& the MiMeS 			Collaboration 2013, ApJ, 764, 171
\bibitem[de Mink et al.(2013)]{demink2013}
	de Mink, S.~E., Langer, N., Izzard, R.~G., Sana, H. \& de Koter, A. 			2013, ApJ, 764, 166
\bibitem[Donati et al.(1992)]{donati1992} 
	Donati, J.-F., Semel, M. \& Rees, D. E. 1992, A\&A, 265, 669
\bibitem[Donati et al.(1997)]{donati1997}
	Donati, J.-F., Semel, M., Carter, B.~D., Rees, D.~E., \& 
	Collier Cameron, A. 1997, MNRAS, 291, 658
\bibitem[Donati et al.(2006)]{donati2006}
	Donati, J.-F., Howarth, I.~D., Jardine, M.~M., et al. 2006, MNRAS, 370, 		629
\bibitem[Donati \& Landstreet(2009)]{donati2009}
	Donati, J.-F. \& Landstreet, J.~D. 2009, ARA\&A, 47, 333
\bibitem[Ferrario et al.(2009)]{ferrario2009}
	Ferrario, L., Pringle, J.~E., Tout, C.~A. \& Wickramasinghe, D.~T. 2009, 	MNRAS, 400, L71
\bibitem[Fitzpatrick(1999)]{fitz1999}
	Fitzpatrick, E.~L. 1999, PASP, 111, 63
\bibitem[Flaccomio et al.(2000)]{flaccomio2000}
	Flaccomio, E., Micela, G., Sciortino, S., et al. 2000, A\&A, 355, 651
\bibitem[Flaccomio et al.(2006)]{flaccomio2006}
	Flaccomio, E., Micela, G. \& Sciortino, S. 2006, A\&A, 455, 903
\bibitem[Fossati et al.(2009)]{fossati2009}
	Fossati, L., Ryabchikova, T., Bagnulo, S., et al. 2009, A\&A, 503, 945
\bibitem[Glagolevskij(1994)]{glago1994}
	Glagolevskij, Y.~V. 1994, Bulletin of the Special Astrophysics 				Observatory, 34, 152
\bibitem[Glebbeek et al.(2013)]{gleb2013}
	Glebbeek, E., Gaburov, E., Portegies Zwart, S. \& Pols, O.~R. 2013, 			MNRAS, 434, 3497
\bibitem[Gray(1965)]{gray1965}
	Gray D.F. 1965, AJ, 70, 362
\bibitem[Grevesse et al.(1996)]{grevesse1996}
        Grevesse, N., Noels, A. \& Sauval, A.~J. 1996, 
        Astronomical Society of the Pacific Conference Series, 99, 117
\bibitem[Grunhut et al.(2008)]{grunhut2013} 
	Grunhut, J.~H., Wade, G.~A., Leutenegger, M., et a. 2013, MNRAS, 428, 			1686
\bibitem[Hareter et al.(2008)]{hareter2008} 
	Hareter, M., Reegen, P., Kuschnig, R., et al. 2008, CoAst, 156, 48
\bibitem[Hauck \& Mermilliod(1997)]{hauck1997}
	Hauck, B. \& Mermilliod, M. 1997, A\&AS, 129, 431
\bibitem[Hempel \& Holweger(2003)]{HH2003}
 	Hempel, M., \& Holweger, H. 2003, A\&A, 408, 1065
\bibitem[Hibbert(1988)]{H}
 	Hibbert, A. 1988, Phys. Scr., 38, 37
\bibitem[Hohle et al.(2010)]{hohle2010}
	Hohle, M.~M., Neuh{\"a}user, R. \& Schutz, B.~F. 2010, AN, 331, 349
\bibitem[Kaiser(2006)]{kaiser2006}
	Kaiser, A. 2006, Determination of Fundamental Parameters with Stroemgren 	Photometry, in Astronomical Society of the Pacific Conference Series, 			eds C. Aerts \& C. Sterken, 349, 257
\bibitem[Kallinger et al.(2008)]{kallinger2008} 
	Kallinger, T., Reegen, P. \& Weiss, W. W. 2008, A\&A, 481, 571
\bibitem[Kharchenko et al.(2004)]{kharchenko2004}
	Kharchenko, N.~V., Piskunov, A.~E., R{\"o}ser, S., Schilbach, E. \& 			Scholz, R.-D. 2004, AN, 325, 740
\bibitem[Kharchenko et al.(2005)]{kharchenko2005}
	Kharchenko, N.~V., Piskunov, A. E., R{\"o}ser, S., Schilback, E. \& 			Scholz, R.-D. 2005, A\&A, 438, 1163
\bibitem[Kochukhov(2007)]{synth3}
	Kochukhov, O. 2007, Spectrum synthesis for magnetic, chemically 
	stratified stellar atmospheres, in Magnetic Stars 2006, eds 
	I.~I.~Romanyuk, D.~O.~Kudryavtsev, O.~M.~Neizvestnaya, 
	\& V.~M.~Shapoval, 109, 118
\bibitem[Kochukhov et al.(2010)]{kochukhov2010} 
	Kochukhov, O., Makaganiuk, V. \& Piskunov, N. 2010, A\&A, 524, A5
\bibitem[Kochukhov et al.(2011)]{kochukhov2011}
	Kochukhov, O., Lundin, A., Romanyuk, I. \& Kudryavtsev, D. 2011, ApJ, 			726, 24
\bibitem[Korntreff et al.(2012)]{korn2012}
	Korntreff, C., Kaczmarek, T. \& Pfalzner, S. 2012, A\&A, 543, A126
\bibitem[Kunzli et al.(1997)]{kunzli1997} 
	Kunzli, M., North, P., Kurucz, R.~L. \& Nicolet, B. 1997, A\&AS, 122, 51
\bibitem[Kupka et al.(1999)]{vald2} 
	Kupka, F., Piskunov, N., Ryabchikova, T. A., Stempels, H. C., Weiss, W. 		W. 1999, A\&AS, 138, 119
\bibitem[Kuschnig et al.(1997)]{kuschnig1997} 
	Kuschnig, R., Weiss, W. W., Gruber, R., Bely, P. Y., Jenkner, H. 1997, 			A\&A, 328, 544
\bibitem[Langer(2012)]{langer2012}
	Langer, N. 2012, A\&ARA, 50, 107
\bibitem[Langer(2013)]{langer2013}
	Langer, N. 2013, Magnetic Fields in Stars: Origin and Impact, in 			proceedings of IAU-Symposium 302: Magnetic fields throughout stellar 			evolution, in press (arXiv: 1312.2373)
\bibitem[Lata(2002)]{lata2002}
	Lata, S., Pandey, A.~K., Sagar, R. \& Mohan, V. 2002, A\&A, 388, 158
\bibitem[Lefever(2007)]{lefever2007}
	Lefever, K. 2007, Ph.D. thesis, K. U. Leuven
\bibitem[Lenz \& Breger(2005)]{lenz2005} Lenz, P. \& Breger, M. 2005, CoAst, 			146, 53
\bibitem[Liu et al.(1991)]{liu1991}
	Liu, T., Janes, K.~A. \& Bania, T.~M. 1991, AJ, 102, 1103
\bibitem[Loktin(2001)]{lotkin2001}
	Loktin, A.~V., Gerasimenko, T.~P. \& Malysheva, L.~K. 2001, Astronomical 	and Astrophysical Transactions, 20, 607
\bibitem[Lynga(1985)]{lynga1987}
	Lunga, G. 1985, Computer-Based Catalogue of Open Cluster Data, IAU 			Symposium, ed. van Woerden, H., Allen, R.~J. \& Burton, W.~B., 106, 143
\bibitem[Maeder \& Meynet(2012)]{maeder2012}
	Maeder, A. \& Meynet, G. 2012, Reviews of Modern Physics, 84, 25
\bibitem[Malysheva(1997)]{malysheva1997}
	Malysheva, L.~K. 1997, Astronomy Letters, 23, 585
\bibitem[Martins et al.(2012)]{martins2012}
	Martins, F., Escolano, C., Wade, G.~A., Donati, J.~F., Bouret, J.~C. \& 		Mimes Collaboration 2012, A\&A, 538, A29
\bibitem[Massey et al.(2013)]{massey2013}
	Massey, P., Neugent, K. F., Hillier, D. J. \& Puls, J. 2013, ApJ, 768, 6
\bibitem[Mermilliod(1991)]{mermilliod1991}
	Mermilliod, J. C. 1991, Catalogue of Homogeneous Means in the UBV 			System, VizieR On-line Data Catalog: II/168
\bibitem[Mermilliod \& Paunzen(2003)]{webda}
	Mermilliod, J.-C., \& Paunzen, E. 2003, A\&A, 410, 511
\bibitem[Michaud(1970)]{michaud1970}
	Michaud, G. 1970, ApJ, 160, 641
\bibitem[Michaud et al.(1979)]{michaud1979}
	Michaud, G., Martel, A., Montmerle, T., Cox, A.~N., Magee, N.~H. \& 			Hodson, S.~W. 1979, ApJ, 234, 206
\bibitem[Moon \& Dworetsky(1985)]{moon1985}
	Moon, T.~T. \& Dworetsky, M.~M. 1985, MNRAS, 217, 305
\bibitem[Morel \& Butler(2008)]{morel08}
 	Morel, T., \& Butler, K. 2008, A\&A, 487, 307
\bibitem[Morel et al.(2013)]{morel2013}
	Morel, T., Briquet, M., Auvergne, M, et al. 2013, A\&A, submitted
\bibitem[Morgan et al.(1965)]{morgan1965}
	Morgan, W.~W., Hiltner, W.~A., Neff, J.~S., Garrison, R. \& Osterbrock, 		D.~E. 1965, ApJ, 142, 974
\bibitem[Moss(2001)]{moss2001}
	Moss, D. 2001, Magnetic Fields in the Ap and Bp Stars: a Theoretical 			Overview, Magnetic Fields Across the Hertzsprung-Russell Diagram, 			Astronomical Society of the Pacific Conference Series, eds. Mathys, G., 		Solanki, S.~K., Wickramasinghe, D.~T., 248, 305
\bibitem[Napiwotzki et al.(1993)]{napiwotzki1993}
	Napiwotzki, R., Schoenberner, D. \& Wenske, V. 1993,
	A\&A, 268, 653
\bibitem[Piskunov et al.(1995)]{vald1} 
	Piskunov, N. E., Kupka, F., Ryabchikova, T. A., Weiss, W. W., \& 			Jeffery, C. S. 1995, A\&AS,  112, 525
\bibitem[Przybilla et al.(2008)]{P08}
	Przybilla, N., Nieva, M.-F., Butler, K. 2008, ApJ, 688, L103
\bibitem[Puls et al.(2005)]{puls2005}
	Puls, J., Urbaneja, M.~A., Venero, R., et al. 2005, A\&A, 435, 669
\bibitem[Puls et al.(2008)]{puls2008}
	Puls, J., Vink, J.~S. \& Najarro, F. 2008, A\&AR, 16, 209
\bibitem[Raskin et al.(2011)]{ras11}
	Raskin, G., van Winckel, H., Hensberge, H., et al., 2011, A\&A 526, 69
\bibitem[Reegen(2007)]{reegen2007} 
	Reegen, P. 2007, A\&A, 467, 135
\bibitem[Renson \& Manfroid(2009)]{renson2009}
	Renson, P. \& Manfroid, J. 2009, A\&A, 498, 961
\bibitem[Rufener et al.(1966)]{rufener1966}
	Rufener, F., Hauck, B., Goy, G., Peytremann, E. \& Maeder, A. 1996, 			Journal des Observateurs, 49, 417
\bibitem[Ryabchikova et al.(1999)]{vald3} 
	Ryabchikova, T.~A., Piskunov, N.~E., Stempels, H.~C., Kupka, F., \& 			Weiss, W.~W. 1999, Phis. Scr., T83, 162
\bibitem[Salpeter(1955)]{salpeter1955}
	Salpeter, E.~E. 1955, ApJ, 121, 161
\bibitem[Sana et al.(2012)]{sana2012}
	Sana, H., de Mink, S.~E., de Koter, A., et al. 2012, Science, 337, 444
\bibitem[Santolaya-Rey et al.(1997)]{sr1997}
	Santolaya-Rey, A.~E., Puls, J. \& Herrero, A. 1997, A\&A, 323, 488
\bibitem[Schaerer \& Schmutz(1994)]{schaerer1994}
	Schaerer, D. \& Schmutz, W. 1994, A\&A, 288, 231
\bibitem[Schneider et al.(2013a)]{schneider2013a}
	Schneider, F.~R.~N., Langer, N., Lau, H.~H.~B. \& Izzard, R.~G. 2013, 			The BONNSAI project: A statistical comparison of individual stars with 			stellar evolution models, Setting a new standard in the analysis of 			binary stars, Leuven, Belgium, in press
\bibitem[Schneider et al.(2013b)]{schneider2013b}
	Schneider, F.~R.~N., Izzard, R.~G., de Mink, S.~E., et al. 2014, ApJ, 			780, 117
\bibitem[Shulyak et al.(2004)]{llm}
        Shulyak, D., Tsymbal, V., Ryabchikova, T., St\"utz\, Ch., \& 
	Weiss, W. W. 2004, A\&A, 428, 993
\bibitem[Sim{\'o}n-D{\'{\i}}az \& Herrero(2007)]{simon2007}
	Sim{\'o}n-D{\'{\i}}az, S. \& Herrero, A. 2007, A\&A, 468, 1063
\bibitem[Sim{\'o}n-D{\'{\i}}az \& Herrero(2007)]{simon2013}
	Sim{\'o}n-D{\'{\i}}az, S. \& Herrero, A. 2013, A\&A, in press (arXiv: 			1311.3360)
\bibitem[Spassova \& Beav(1985)]{spassova1985}
	Spassova N.~M. \& Beav P.~V. 1985, Astrophysics and Space Science, 112, 		111
\bibitem[Sung et al.(1997)]{sung1997}
	Sung, H., Bessel, M. S. \& Lee, S.-W. 1997, AJ, 114, 2644
\bibitem[Sung et al.(2009)]{sung2009}
 	Sung, H., Sauffer, J. R. \& Bessel, M. S. 2009, AJ, 138, 1116
\bibitem[Thompson et al.(1978)]{thompson1978}
	Thompson G.~I., Nandy K., Jamar C., et al. 1978, Catalogue of stellar 			ultraviolet fluxes (TD1): A compilation of absolute stellar fluxes 			measured by the Sky Survey Telescope (S2/68) aboard the ESRO satellite 			TD-1, VizieR On-line Data Catalog: II/59B
\bibitem[van der Bliek et al.(1996)]{bliek1996} 
	van der Bliek, N. S., Manfroid, J. \& Bouchet, P. 1996, A\&AS, 119, 547
\bibitem[Vick et al.(2010)]{vick2010} 
	Vick, M., Michaud, G., Richer, J. \& Richard, O. 2010, A\&A, 521, A62
\bibitem[Vogel \& Kuhi(1981)]{vogel1981}
	Vogel, S.~N. \& Kuhi, L.~V. 1981, ApJ, 245, 960 
\bibitem[Wade et al.(2012)]{mimes}
	Wade, G.~A., Grunhut, J.~H. \& MiMeS Collaboration 2012, The MiMeS 			Survey of Magnetism in Massive Stars in Circumstellar Dynamics at High 			Resolution, Astronomical Society of the Pacific Conference Series, eds 			Carciofi, A.~C. \& Rivinius, T., 464, 405 
\bibitem[Wade et al.(2013)]{wade2013}
	Wade, G.~A., Grunhut, J., Alecian, E., et al. 2013, The magnetic 			characteristics of Galactic OB stars from the MiMeS survey of magnetism 		in massive stars, Proceedings of IAUS 302: Magnetic fields throughout 			stellar evolution, in press (arXiv: 1310.3965)
\bibitem[Walker(1956)]{walker1956} 
	Walker, M.~F. 1956, ApJS, 2, 365
\bibitem[Walker et al.(2003)]{walker2003} 
	Walker, G., Matthews, J. M., Kuschnig, R., et al. 2003, PASP, 115, 1023
\bibitem[Weisskopf et al.(2002)]{weisskopf2002} 
	Weisskopf, M. C., Brinkman, B., Canizares, S., et al. 2002, PASP, 114, 1
\bibitem[Werner et al.(2004)]{werner2004} 
	Werner, M. W., Roellig, T. L., Low, F. J., et al. 2004, ApJS, 154, 1
\bibitem[Wright et al.(2010)]{wright2010} 
	Wright, E. L., Eisenhardt, P. R. M., Mainzer, A. K., et al. 2010, AJ, 			140, 1868
\bibitem[Zacharias et al.(2005)]{zacharias2005} 
	Zacharias, N., Monet, D. G., Levine, S. E., et al. 2005, BAAS, 36, 1418
\bibitem[Zwintz et al.(2009)]{zwintz2009}
	Zwintz, K., Hareter, M., Kuschnig, R., et al. 2009, A\&A, 502, 239
\bibitem[Zwintz et al.(2013a)]{zwintz2013a}
 	Zwintz, K., Fossati, L., Ryabchikova, T., et al. 2013, A\&A, 550, A121
\bibitem[Zwintz et al.(2013b)]{zwintz2013b} 
 	Zwintz, K., Fossati, L., Guenther, D.~B., et al. 2013, A\&A, 552, A68
\end{thebibliography}
\end{document}